\documentclass[amsmath,amssymb,aps,prd, nofootinbib, superscriptaddress]{revtex4}%{elsarticle}
\usepackage{graphicx}
\usepackage{dcolumn}
\usepackage{bm}
\usepackage[usenames,dvipsnames]{xcolor}
\usepackage[colorlinks=true,linkcolor=NavyBlue,citecolor=NavyBlue,urlcolor=Blue]{hyperref}
\usepackage[mathlines]{lineno}
\usepackage{amsfonts,color}
\usepackage{tensor}

\begin{document}
	
	\title{The Gauss-Bonnet topological scalar in the Geometric Trinity of Gravity}

 \author{Francesco Bajardi}
	\altaffiliation[]{\href{f.bajardi@ssmeridionale.it}{f.bajardi@ssmeridionale.it}}
	\affiliation{Scuola Superiore Merdionale, Largo S. Marcellino 10, I-80138, Napoli, Italy}
 \affiliation{Istituto Nazionale di Fisica Nucleare (INFN),  sez. di Napoli, Complesso Universitario di Monte S. Angelo, via Cinthia, Ed. N, 80126 Napoli, Italy}
	\author{Daniel Blixt}
	\altaffiliation[]{\href{d.blixt@ssmeridionale.it}{d.blixt@ssmeridionale.it}}
	\affiliation{Scuola Superiore Merdionale, Largo S. Marcellino 10, I-80138, Napoli, Italy}
	%\affiliation{Laboratory of Theoretical Physics, Institute of Physics, University of Tartu, W. Ostwaldi 1, Tartu 50411, Estonia}
	
	\author{Salvatore Capozziello}
	\altaffiliation[]{\href{capozziello@unina.it}{capozziello@unina.it}}
	\affiliation{Scuola Superiore Merdionale, Largo S. Marcellino 10, I-80138, Napoli, Italy}
 \affiliation{Istituto Nazionale di Fisica Nucleare (INFN),  sez. di Napoli, Complesso Universitario di Monte S. Angelo, via Cinthia, Ed. N, 80126 Napoli, Italy}
	\affiliation{Dipartimento di Fisica "E. Pancini", Universit\'a di Napoli Federico II, Complesso Universitario di Monte S. Angelo, via Cinthia, Ed. N, 80126 Napoli, Italy}

	\date{\today}
	
	\begin{abstract}
		The Gauss-Bonnet topological scalar is presented in metric-teleparallel formalism as well as in the symmetric and general teleparallel formulations. In all of the aforementioned frameworks, the full expressions are provided explicitly in terms of torsion, non-metricity and Levi-Civita covariant derivative. The number of invariant terms of this form is counted and compared with the number which can appear in the corresponding effective field theory. Although the difference in this number is not very large, it is found that the Gauss-Bonnet invariant excludes some of the effective field theory terms. This result sheds new light on how General Relativity symmetries can be maintained at higher order in teleparallel theories: this fact  appears to be highly nontrivial in the teleparallel formulation. The importance of the so-called ``pseudo-invariant'' theories like $f(T)$- and $f(T,T_\mathcal{G})$-gravity is further discussed in the context of teleparallel Gauss-Bonnet gravity.
	\end{abstract}
	
	\maketitle
	
	\section{Introduction}
	General relativity (GR) has successfully predicted several observations that were not explained by Newtonian mechanics. In particular, GR fits very well the observational data at Solar System scales and  predicted also astrophysical objects and processes such as black holes and gravitational waves. At cosmological scales, the Newtonian limit can often be taken. However, it is important to include dark energy and dark matter for consistency with observations. For this reason, the standard model of cosmology, the so-called $\Lambda$CDM-model, includes dark side sources. Here $\Lambda$ is the cosmological constant, which is predicted by GR and CDM stands for cold dark matter (modeled by particle physics). Cosmology is typically considered at early  and late time. The reason  is that we cannot trace a comprehensive cosmic history ranging from  the very beginning of the universe up to today epoch.  There is observational evidence that during early times the universe went through a period of accelerated expansion, called inflation. On the other hand,   another accelerated expansion is working today after the cosmic flow passed through a period of structure formation. No cosmological model, up to date,  succeeded in giving a self-consistent picture valid at any time, that is capable of representing the universe at UV and IR scales,  despite of the fact that observational data are precise and reliable (i.e. thanks to the so-called {\it Precision Cosmology}).
		
	Although, the observational success of GR does not come completely without any shortcomings. In contrast to the standard model of particle physics, GR as a theory is not renormalizable. This has the consequence that the theory is not UV-complete and therefore we cannot be certain of how to deal with predictions at scales where both gravity and quantum physics become relevant. However, it is possible to quantize GR at orders of magnitude smaller than the scale at which it breaks down and, in this context, GR is treated as an effective field theory (EFT). In this perspective, it is expected that higher order curvature terms, beyond the standard Ricci scalar of the Einstein-Hilbert (EH) action,  become relevant at certain scales. Taking these effects into account, for example during the inflation,  leads to predictions that are consistent with observations \cite{Planck:2018jri}. 
	
	Other shortcomings of GR  include the observational value of the cosmological constant, which is much smaller than the predicted one at early epochs, and the Hubble constant tension \cite{Planck:2018vyg}. In addition to theoretical puzzles, these are some of the reasons for the growing interest in examining modifications and extensions of Einstein's theory. Most of them, generalize GR by extending the gravitational action, including \emph{e.g.} functions of the scalar curvature, higher-order curvature terms or the coupling with dynamical scalar fields \cite{Capozziello:2011et}. In particular, theories including quadratic curvature terms are  interesting because  motivated by GR as an EFT for  constructing effective models towards quantum gravity. In particular, it was found that a certain combination of higher-order contractions of the Riemann tensor, is a topological surface in four dimensions. Such a topological surface is the {\it Gauss-Bonnet topological scalar} and, in four dimensions, turns out to be the Euler density which, according to the generalized Gauss-Bonnet theorem, once integrated over the manifold provides the Euler characteristic. The Gauss--Bonnet invariant $\mathcal{G}$ is often considered due to its topological nature, as a tool to reduce the dynamics; however, in order to make its contribution not-trivial in four dimensions, it is usually coupled to a dynamical scalar field \cite{Carter:2005fu, Bajardi:2020xfj} or, sometimes, a function of it is taken into account into the gravitational action \cite{Cognola:2006eg}. In the latter approach, the function $f(\mathcal{G})$ can be added to the scalar curvature in order to mimic the cosmological constant at late times \cite{Li:2007jm, Bajardi:2022tzn} and, when $f(\mathcal{G}) = 0$, GR can be safely recovered. A further approach is to deal with higher-dimensions (where $\mathcal{G}$ is not-trivial) and couple the Gauss-Bonnet invariant to a constant that diverges when $D = 4$ \cite{Gurses:2020rxb, Glavan:2019inb}. Generally, modified theories somehow dealing with the Gauss--Bonnet topological term have been deeply investigated in recent years, as well as applications to cosmology and astrophysics \cite{Benetti:2018zhv, SantosDaCosta:2018bbw, Chakravarti:2022zeq, Oikonomou:2015qha, Odintsov:2020sqy, Odintsov:2018nch}.

 Extending the gravitational action is not the only way to modify GR, as several other assumptions of the theory can be relaxed \cite{Cai:2015emx}. For instance, it is possible to consider Christoffel connections that are not metric-compatible, where both torsion and non-metricity arise in the given space-time. As the former is linked to the anti-symmetric part of the connection, the latter is related to the fact that  covariant derivative of the metric tensor can be non-vanishing. In this framework, the Riemann tensor of GR and its contractions can be rewritten in terms of torsion, non-metricity, or both. The three theories, described by curvature, torsion, and non-metricity, are dynamically equivalent as the corresponding actions only differ for a boundary term. This is typically referred as the {\it geometric trinity of gravity} \cite{BeltranJimenez:2019esp}. In particular, this is viewed as a classical equivalence with boundary terms, which do not contribute to the field equations and are dropped in their respective action formulation (see more details about the different formulations in section \ref{sec:Trinity}). From the GR point of view, the trinity of gravity is motivated mainly by the possibility to formulate GR as a gauge theory \cite{Capozziello:2022zzh,Krssak:2018ywd} and to predict a well-defined mass in gravity opposed to the EH action formulation \cite{Gomes:2022vrc}. From a modified gravity point of view, formulating gravity in terms of torsion and non-metricity allows violation of Lorentz and diffeomorphism invariance, respectively. 
	
	This article focuses on quadratic gravity and, in particular, on the formulation of the Gauss-Bonnet invariant in the trinity of gravity. In Sec. \ref{sec:Trinity} it is discussed  how the trinity of gravity is realized, while, in Sec. \ref{sec:GB}, we demonstrate how the Gauss-Bonnet invariant can be obtained in the trinity of gravity. Furthermore, this is compared with the EFT point of view, which was recently considered in \cite{Baldazzi:2021kaf} and \cite{Mylova:2022ljr}. In Sec. \ref{sec:Principles}, starting from the previous results, we discuss some fundamental principles for the construction of teleparallel theories of gravity.  Finally, Sec. \ref{sec:Conclusions} is devoted to conclusions and future perspectives. Appendix \ref{app:GB} contains explicit expressions to support the tables in Sec. \ref{sec:GB}.
	
	\section{The Einstein-Hilbert action and its equivalences}
	\label{sec:Trinity}
	When considering curvature defined by a general linear affine connection $\Gamma^\rho{}_{\mu\nu}$, instead of  the Levi-Civita connection, the Riemann tensor can be generally written as:
	\begin{align}
	\begin{split}
	R^\mu{}_{\nu \rho \sigma }(\Gamma)=2\partial_{[\rho}\Gamma^\mu{}_{|\nu|\sigma]}+2\Gamma^\mu{}_{\lambda[\rho}\Gamma^\lambda{}_{|\nu|\sigma]}.
	\end{split}
	\end{align}
	In metric formalism,  $\Gamma^\rho{}_{\mu\nu}$ is taken to be the Levi-Civita connection, which is be denoted as
	\begin{align}
	\begin{split}
	\begin{Bmatrix}
	\rho \\ \mu \nu 
	\end{Bmatrix}=\frac{1}{2}g^{\alpha\lambda}\left( g_{\lambda\nu ,\mu}+g_{\mu\lambda ,\nu} -g_{\mu\nu , \lambda} \right).
	\end{split}
	\end{align}
	The part of the connection which is not metric compatible is defined to be the so-called non-metricity, namely
	\begin{align}
	Q_{\rho\mu\nu}:=\nabla_\rho g_{\mu\nu},
	\end{align}
	where $\nabla$ is the covariant derivative with respect to $\Gamma$, whose antisymmetric part is defined as the torsion
	\begin{align}
	T^\rho{}_{\mu\nu}:=2\Gamma^\rho{}_{[\mu\nu]}.
	\end{align}
	Therefore, the linear affine connection $\Gamma^\rho{}_{\mu\nu}$ can be decomposed in terms of torsion, non-metricity and metric-compatible, as
	\begin{align}
	\Gamma^\alpha{}_{\mu\nu}=\begin{Bmatrix}
	\alpha \\ \mu \nu
	\end{Bmatrix}+K^\alpha{}_{\mu\nu}\left(T \right)+L^\alpha{}_{\mu\nu}\left(Q\right),
	\end{align}
	where the two additional terms arising in this formulation, namely $\tensor{K}{^\alpha_\mu_\nu}$ and $\tensor{L}{^\alpha_\mu_\nu}$, are related to torsion and non-metricity, respectively. The former is called contortion tensor and is defined as:
	\begin{align}
	\tensor{K}{^\alpha_\mu_\nu}(T) = -\frac{1}{2}g^{\alpha\beta}\left(T_{\beta\mu\nu}+ T_{\nu\mu\beta} + T_{\mu\nu\beta} \right),
	\end{align}
	while the latter is the disformation tensor, that is
	\begin{align}
	\tensor{L}{^\alpha_\mu_\nu}(Q) = \frac{1}{2}g^{\alpha\beta}\left( Q_{\beta\mu\nu} - Q_{\mu\nu\beta} - Q_{\nu\beta\mu} \right) .
	\end{align}
	Through these relations, the curvature of a general affine connection becomes
	\begin{align}
	\begin{split}
	\label{eq:Trinity}
	R^\mu{}_{\nu\rho\sigma}(\Gamma)= & \tensor{R}{^\mu_\nu_\rho_\sigma}\left( \{\} \right) + 2\mathcal{D}_{[\rho}\tensor{K}{^\mu_|_\nu_|_\sigma_]} + 2 \tensor{K}{^\mu_\lambda_[_\rho}\tensor{K}{^\lambda_|_\nu_|_\sigma_]}+  2\mathcal{D}_{[\rho}\tensor{L}{^\mu_|_\nu_|_\sigma_]} + 2 \tensor{L}{^\mu_\lambda_[_\rho}\tensor{L}{^\lambda_|_\nu_|_\sigma_]} + \\
	& 2\tensor{K}{^\mu_\lambda_[_\rho}\tensor{L}{^\lambda_|_\nu_|_\sigma_]} + 2\tensor{L}{^\mu_\lambda_[_\rho}\tensor{K}{^\lambda_|_\nu_|_\sigma_]},
	\end{split}
	\end{align}
	where $\tensor{R}{^\mu_\nu_\rho_\sigma}\left( \{\} \right)$ is the Riemann tensor for the Levi-Civita connection.
	It follows that the Ricci scalar of $\Gamma$ is
	\begin{align}
	\begin{split}
	\label{eq:Ricci}
	R(\Gamma)&=g^{\nu\sigma}R^\mu{}_{\nu\mu\sigma}(\Gamma)=\mathcal{R}+\mathbb{T}(T)+\mathbb{Q}(Q)+Q_{\mu\nu\rho}T^{\rho\mu\nu}+Q_\mu T^\mu -\Bar{Q}_\mu T^\mu\\
	&+\mathcal{D}_\mu \left(Q^\mu -\Bar{Q}^\mu -2T^\mu \right),
	\end{split}
	\end{align}
	where $\mathcal{R}$ is the Ricci scalar of the Levi-Civita connection, $\mathcal{D}$ is the covariant derivative with respect to the Levi-Civita connection, $T_\mu=T^\rho{}_{\rho\mu}, \ Q_\mu=Q_{\mu\rho}{}^\rho$, and $\Bar{Q}_\mu=Q^\rho{}_{\rho\mu}$. Teleparallelism is defined by $R^\mu{}_{\nu\rho\sigma}(\Gamma)$. It is then evident that the EH action can be rewritten (using Eq. \eqref{eq:Ricci}) in the following way
	\begin{align}
	\begin{split}
	S_\mathrm{EH}&=\frac{1}{2\kappa}\int \mathrm{d}^4 x \sqrt{-g} \mathcal{R}\\
	&=-\frac{1}{2\kappa}\int \mathrm{d}^4 x \sqrt{-g}\left( \mathbb{T}+\mathbb{Q}
	+Q_{\mu\nu\rho}T^{\rho\mu\nu} \right. \left.+Q_\mu T^\mu -\Bar{Q}_\mu T^\mu
	+\mathcal{D}_\mu \left(Q^\mu -\Bar{Q}^\mu -2T^\mu \right)\right)\\
	&:= S_\mathrm{GTEGR}-\frac{1}{2\kappa}\int \mathrm{d}^4 x \sqrt{-g}\mathcal{D}_\mu\left(Q^\mu-\Bar{Q}^\mu -2T^\mu \right).
	\end{split}
	\end{align}
	Here, the last term becomes a boundary term and $S_\mathrm{GTEGR}$ is called the ``General Teleparallel Equivalent to General Relativity'' since its field equations correspond to those of GR, as they are unaffected by the boundary term. Furthermore, metric teleparallelism is defined with the additional condition that the nonmetricity tensor $Q_{\alpha\mu\nu}$ vanishes. This special case transforms $S_\mathrm{GTEGR}$ to $S_\mathrm{TEGR}$, which in turn is called the ``Teleparallel Equivalent to General Relativity''\footnote{For historical reasons teleparallel theories assumed nonmetricity to be zero. A better name could be ``Metric Teleparallel Equivalent to General Relativity.}
	\begin{align}
	\begin{split}
	S_\mathrm{EH}&=\frac{1}{2\kappa}\int \mathrm{d}^4 x \sqrt{-g} \mathcal{R}=-\frac{1}{2\kappa}\int \mathrm{d}^4 x \sqrt{-g}\left( \mathbb{T}
	-2\mathcal{D}_\mu T^\mu\right)\\
	&:= S_\mathrm{TEGR}+\frac{1}{\kappa}\int \mathrm{d}^4 x \sqrt{-g}\mathcal{D}_\mu T^\mu .
	\end{split}
	\end{align}
	Symmetric teleparallelism, on the other hand, assumes that the connection $\Gamma$ is symmetric with respect to the lowest indexes, which implies vanishing torsion. In this limit, $S_\mathrm{GTEGR}$ becomes the ``Symmetric Teleparallel Equivalent to General Relativity'': 
	\begin{align}
	\begin{split}
	S_\mathrm{EH}&=\frac{1}{2\kappa}\int \mathrm{d}^4 x \sqrt{-g} \mathcal{R}=-\frac{1}{2\kappa}\int \mathrm{d}^4 x \sqrt{-g}\left( \mathbb{Q}
	+\mathcal{D}_\mu \left(Q^\mu -\Bar{Q}^\mu  \right)\right)\\
	&:= S_\mathrm{STEGR}-\frac{1}{2\kappa}\int \mathrm{d}^4 x \sqrt{-g}\mathcal{D}_\mu\left(Q^\mu-\Bar{Q}^\mu \right).
	\end{split}
	\end{align}
    	Clearly, all of these theories have the same field equations satisfying the same symmetries as GR and they are physically completely indistinguishable. However, they give an alternative framework for working on modified theories of gravity. By allowing for torsion and non-metricity, it is possible to construct more invariants than those found in purely Riemannian geometries. Furthermore, for nonlinear extensions like $f(T)$- and $f(Q)$-gravity \cite{Bajardi:2023vcc, Bajardi:2021tul, BeltranJimenez:2019tme, Khyllep:2021pcu, Capozziello:2019cav, Cai:2019bdh}, the boundary term plays a nontrivial role. In this article we will focus on quadratic teleparallel theories of gravity, that is, teleparallel models containing fourth-order invariants into the action. This have already been done from an EFT point of view in  \cite{Baldazzi:2021kaf} and, for the metric teleparallel case, this was studied by including dynamical scalar fields \cite{Mylova:2022ljr}. Specifically, the main focus will be on how the Gauss-Bonnet invariant, which in four dimensions is purely topological, appears in teleparallel quadratic gravity. Another reason for exploring the Gauss-Bonnet invariant is that the calculations involve quadratic contractions of the Riemann tensor (and Ricci tensor). This gives an insight on how to preserve or break the symmetries that the Riemann tensor satisfies in the construction of modified teleparallel theories of gravity. Applications of these considerations can be particularly important in cosmology \cite{Capozziello:2022tvv}.	
    	
	\section{The Gauss-Bonnet invariant in the geometric  trinity of gravity}
	\label{sec:GB}
The Gauss-Bonnet invariant is usually defined as the exterior product between \textbf{R} $\wedge$ \textbf{R}, with \textbf{R} being the two-form curvature. It naturally arises when considering higher dimensions (\emph{e.g.} in Lovelock or Born-Infield theories), because,  in four dimensions, it reduces to a topological surface term which does not contribute to the field equations and its integration over the manifold provides the so called Euler Characteristic. In coordinate representation, the Gauss-Bonnet invariant can be expressed as:
	\begin{align}
 \label{GBcoord}
	\mathcal{G}=R^2-4 R^{\mu\nu}R_{\mu\nu}+R^{\mu\nu\rho\sigma}R_{\mu\nu\rho\sigma}.
	\end{align}
 As mentioned in the introduction, models containing the Gauss-Bonnet invariant caught particular interest in several contexts, due to the capability of $\mathcal{G}$ to simplify the dynamics of the system and provide an alternative explanation for dark energy and other shortcomings occurring in GR \cite{Nojiri:2005jg, Cognola:2006eg}. In particular, the introduction of Gauss-Bonnet invariant into the gravitational action can contribute to give ghost-free models \cite{Nojiri:2021mxf, Nojiri:2019dwl}. Recent works have also considered modified actions containing a function of both $R$ and $\mathcal{G}$, showing that this theory can predict different phases of the universe evolution without any cosmological constant \cite{Li:2007jm, Elizalde:2010jx, DeLaurentis:2015fea, deMartino:2020yhq}. Similarly, other theories start from actions where a function of $\mathcal{G}$ is added to the scalar curvature, with the consequence that $f(\mathcal{G})$ can play the role of an effective cosmological constant \cite{Zhou:2009cy, DeFelice:2009aj}. Another possibility is  assuming  gravity to be described only by some function of the Gauss--Bonnet invariant, without imposing the GR limit as a requirement \cite{Bajardi:2020osh, Bajardi:2019zzs}. The latter approach, although is still to be investigated, showed that Einstein's gravity can be recovered at the level of equations in some cosmological frameworks. When higher-dimensions are considered, the Gauss-Bonnet invariant becomes not-trivial and is naturally included into the gravitational action as predicted by the Lovelock hypothesis \cite{Lovelock:1971yv, Mardones:1990qc, Exirifard:2007da, Bajardi:2021hya}. Some sub-classes of the Lovelock Lagrangian including the Gauss-Bonnet invariant, are also \emph{quasi}-gauge invariant, namely invariant under gauge transformations up to a boundary term. This is the case \emph{e.g.} of $AdS_5$ Chern-Simons gravity \cite{Miskovic:2007mg, Miskovic:2010ui, Kofinas:2008ub}.
	Our purpose is to express Eq. \eqref{GBcoord} in terms of torsion and nonmetricity, when the condition of teleparallelism is satisfied. The explicit expressions are, indeed, lengthy and for this reason they are presented in Appendix \ref{app:GB}. In what follows, we will start by writing them in terms of contortion and disformation tensors and then we define objects that have characteristic differences. 
	
	\subsection{Metric teleparallel Gauss-Bonnet invariant}
	Let us now consider the teleparallel equivalent of Eq. \eqref{GBcoord}, recast in terms of the contortion tensor. Assuming vanishing non-metricity, the Gauss-Bonnet invariant can be written as
	\begin{align}
	\label{eq:TeleGB}
	\begin{split}
	\mathcal{G}=\mathcal{G}_{\mathrm{TEGR}}=\mathcal{G}_T+\mathcal{G}_{\mathcal{DT}}+\mathcal{G}_{\mathcal{D}\mathcal{D}T},
	\end{split}
	\end{align}
	where 
	\begin{align}
	\begin{split}
	\mathcal{G}_T&:=4K^\kappa{}_{\lambda[\kappa}K^\lambda{}_{|\mu|\nu]}K^\alpha{}_{\beta[\alpha}K^\beta{}_{|\rho|\sigma]}g^{\mu\nu}g^{\rho\sigma}-16K^\rho{}_{\sigma[\rho}K^\sigma{}_{|\mu|\nu]}K_\kappa{}^{\lambda[\kappa}K_\lambda{}^{|\mu|\nu]}\\
	&+4K^\mu{}_{\lambda[\rho}K^\lambda{}_{|\nu|\sigma]}K_{\mu}{}^{\kappa[\rho}K_{\kappa}{}^{|\nu|\sigma]}\\
	&=\mathbb{T}^2-16K^\rho{}_{\sigma[\rho}K^\sigma{}_{|\mu|\nu]}K_\kappa{}^{\lambda[\kappa}K_\lambda{}^{|\mu|\nu]}+4K^\mu{}_{\lambda[\rho}K^\lambda{}_{|\nu|\sigma]}K_{\mu}{}^{\kappa[\rho}K_{\kappa}{}^{|\nu|\sigma]},
	\end{split}
	\end{align}
	\begin{align}
	\begin{split}
	\mathcal{G}_{\mathcal{D}T}&:=8\mathcal{D}_{[\kappa}K^\kappa{}_{|\mu|\nu]}K^\alpha{}_{\beta[\alpha}K^\beta{}_{|\rho|\sigma]}g^{\mu\nu}g^{\rho\sigma}-32\mathcal{D}_{[\kappa}K^\kappa{}_{|\mu|\nu]}K_\rho{}^{\sigma[\rho}K_\sigma{}^{|\mu|\nu]}\\
	&+8\mathcal{D}_{[\rho}K^\mu{}_{|\nu|\sigma]}K_{\mu}{}^{\kappa[\rho}K_{\kappa}{}^{|\nu|\sigma]}\\
	&=-4\mathbb{T}\mathcal{D}_{[\kappa}K^\kappa{}_{|\mu|\nu]}g^{\mu\nu}-32\mathcal{D}_{[\kappa}K^\kappa{}_{|\mu|\nu]}K_\rho{}^{\sigma[\rho}K_\sigma{}^{|\mu|\nu]}+8\mathcal{D}_{[\rho}K^\mu{}_{|\nu|\sigma]}K_{\mu}{}^{\kappa[\rho}K_{\kappa}{}^{|\nu|\sigma]},
	\end{split}
	\end{align}
	\begin{align}
	\begin{split}
	\mathcal{G}_{\mathcal{D}\mathcal{D}T}&:=4\mathcal{D}_{[\kappa}K^\kappa{}_{|\mu|\nu]}\mathcal{D}_{[\lambda}K^\lambda{}_{|\rho|\sigma]}g^{\mu\nu}g^{\rho\sigma}-16\mathcal{D}_{[\rho}K^\rho{}_{|\mu|\nu]}\mathcal{D}^{[\sigma}K_\sigma{}^{|\mu|\nu]}\\
	&+4\mathcal{D}_{[\rho}K^\mu{}_{|\nu|\sigma]}\mathcal{D}^{[\rho}K_\mu{}^{|\nu|\sigma]}.
	\end{split}
	\end{align}
	It is worth pointing out that a lot of attention in the literature has been also directed to another metric teleparallel reformulation of the Gauss-Bonnet invariant, that is \cite{Kofinas:2014owa,Bahamonde:2016kba,Capozziello:2016eaz}
	\begin{align}
	\mathcal{G}=-T_\mathcal{G}+B_\mathcal{G},
	\end{align}
	where \cite{Bahamonde:2016kba}
	\begin{align}
	\label{eq:TG}
	\begin{split}
	T_\mathcal{G}&:= \left(K_a{}^i{}_e K_b{}^ej K_c{}^k{}_f K_d{}^{fl}+2K_a{}^{ij}K_b{}^k{}_e K_f{}^{el}K_d{}^f{}_c+2\partial_d K_a{}^{ij}K_b{}^k{}_{c}{}^{el} \right)\delta^{abcd}_{ijkl}
	\end{split}
	\end{align}
	and
	\begin{align}
	\begin{split}
	B_\mathcal{G}&:= \frac{1}{\sqrt{-g}}\delta^{abcd}_{ijkl}\partial_a\left( \frac{1}{2}K_b{}^{ij}R^{kl}{}_{cd}+K_b{}^{ij}K_c{}^k{}_f K_d{}^{fl} \right).
	\end{split}
	\end{align}
	The advantage of $T_\mathcal{G}$ is that the order of derivatives is reduced by one order compared to $\mathcal{G}_T$, which can make some calculations simpler. However, it is important to note that $T_\mathcal{G}$ is not written in the first-order formalism, in contrast to the standard parallelism of the torsion scalar, which can be motivated to be preferred over Ricci scalar, since the latter contains second order derivatives making TEGR more similar to the standard model of particle physics. 
	
	Furthermore, we would like to stress that recasting the Gauss-Bonnet invariant to $T_\mathcal{G}$ lacks some theoretical motivations. This boundary-term spoils the Lorentz invariance of type II \cite{Bahamonde:2016kba} (as defined in \cite{Blixt:2022rpl}), whereas the field equations are Lorentz invariant. This kind of action formulations is called Lorentz pseudo-invariant \cite{Ferraro:2020tqk,Blixt:2022rpl}. The general take-home lesson from $f(\mathbb{T})$-gravity is that when adding a boundary term and then considering nonlinear extensions, the same number of degrees of freedom occurs at lowest order of perturbations around the most simplest background \cite{BeltranJimenez:2020fvy, BeltranJimenez:2021auj}. However, since the nonlinear extensions (which have been the main topic of applications in the form of $f(T,T_\mathcal{G})$-theories of gravity \cite{Bahamonde:2021gfp}) do not preserve the symmetries of the field equations \cite{Li:2010cg}, they are propagating additional degrees of freedom when taking into account the full nonlinear theory (see \cite{Blixt:2020ekl} and references therein for a review). This insight suggests that there are fields in the theory that suffer from strong coupling and perturbation theory cannot be trusted around this background \cite{BeltranJimenez:2020fvy}. For this reason, in this article we focus on the expression provided by Eq. \eqref{eq:TeleGB}, rather than Eq. \eqref{eq:TG}. Expanding the contortion tensor and writing Eq. \eqref{eq:TeleGB} explicitly in terms of torsion, reveals many invariant teleparallel scalars at quadratic level. The explicit result is very long and therefore is presented in Appendix \ref{app:GB}. Most of the terms occurring in \cite{Baldazzi:2021kaf}, where metric-affine effective field theories are discussed, turns out to be needed in the construction of the metric teleparallel formulation of the Gauss-Bonnet invariant. We list in Table \ref{teletable} a comparison between the number of torsion scalars that appears when expanding the Gauss-Bonnet invariant and the number of all scalars that was found in \cite{Baldazzi:2021kaf}.
	\begin{table}[h]
		\centering
		\begin{tabular}{|c|c|c|c|c|c|c|c|}
			\hline
			$\mathcal{G}_T$ & EFT $T^4$ & $\mathcal{G}_{\mathcal{D}T}$ & EFT $T^2 \mathcal{D}T$ & $\mathcal{G}_{\mathcal{DD}T}$ & EFT $ (\mathcal{D}T)^2$ & Total $\mathcal{G}_\mathrm{TEGR}$ & Total EFT  \\ \hline
			27  & 33 & 18 & 46 & 11 & 16 & 56 & 95  \\ \hline
		\end{tabular}
		\caption{Number of terms involved in the construction of the teleparallel Gauss-Bonnet invariant and number of terms in teleparallel EFT.}
		\label{teletable}
	\end{table}

	\subsection{Symmetric teleparallel Gauss-Bonnet invariant}
	Let us  start again  from Eq. \eqref{GBcoord} but  consider only contributions provided by non-metricity. In this way, $\mathcal{G}$ can be expressed in terms of disformation tensor only. Assuming thus vanishing torsion, the Gauss-Bonnet invariant can be written as
	\begin{align}
\label{GBSTEGR}
	\begin{split}
	\mathcal{G}&:=\mathcal{G}_{\mathrm{STEGR}}=\mathcal{G}_{\mathrm{GTEGR}}=\mathcal{G}_Q+\mathcal{G}_{\mathcal{D}Q}+\mathcal{G}_{\mathcal{D}\mathcal{D}Q},
	\end{split}
	\end{align}
	where
	\begin{align}
	\begin{split}
	\mathcal{G}_Q&:=4L^\kappa{}_{\lambda[\kappa}L^\lambda{}_{|\mu|\nu]}L^\alpha{}_{\beta[\alpha}L^\beta{}_{|\rho|\sigma]}g^{\mu\nu}g^{\rho\sigma}-16L^\rho{}_{\sigma[\rho}L^\sigma{}_{|\mu|\nu]}L_\kappa{}^{\lambda[\kappa}L_\lambda{}^{|\mu|\nu]}\\
	&+4L^\mu{}_{\lambda[\rho}L^\lambda{}_{|\nu|\sigma]}L_\mu{}^{\kappa[\rho}L_\kappa{}^{|\nu|\sigma]}\\
	&=\mathbb{Q}^2-16L^\rho{}_{\sigma[\rho}L^\sigma{}_{|\mu|\nu]}L_\kappa{}^{\lambda[\kappa}L_\lambda{}^{|\mu|\nu]}+4L^\mu{}_{\lambda[\rho}L^\lambda{}_{|\nu|\sigma]}L_\mu{}^{\kappa[\rho}L_\kappa{}^{|\nu|\sigma]},
	\end{split}
	\end{align}
	\begin{align}
	\begin{split}
	\mathcal{G}_{\mathcal{D}Q}&:=8\mathcal{D}_{[\kappa}L^\kappa{}_{|\mu|\nu]}L^\alpha{}_{\beta[\alpha}L^\beta{}_{|\rho|\sigma]}g^{\mu\nu}g^{\rho\sigma}-32\mathcal{D}_{[\rho}L^\rho{}_{|\mu|\nu]}L_\kappa{}^{\lambda[\kappa}L_\lambda{}^{|\mu|\nu]}\\
	&+8\mathcal{D}_{[\rho}L^\mu{}_{|\nu|\sigma]}L_\mu{}^{\kappa[\rho}L_\kappa{}^{|\nu|\sigma]}\\
	&=-4\mathbb{Q}\mathcal{D}_{[\kappa}L^\kappa{}_{|\mu|\nu]}g^{\mu\nu}-32\mathcal{D}_{[\rho}L^\rho{}_{|\mu|\nu]}L_\kappa{}^{\lambda[\kappa}L_\lambda{}^{|\mu|\nu]}+8\mathcal{D}_{[\rho}L^\mu{}_{|\nu|\sigma]}L_\mu{}^{\kappa[\rho}L_\kappa{}^{|\nu|\sigma]},
	\end{split}
	\end{align}
	\begin{align}
	\begin{split}
	\mathcal{G}_{\mathcal{D}\mathcal{D}Q}&:=4\mathcal{D}_{[\kappa}L^\kappa{}_{|\mu|\nu]}\mathcal{D}_{[\lambda}L^\lambda{}_{|\rho|\sigma]}g^{\mu\nu}g^{\rho\sigma}-16\mathcal{D}_{[\rho}L^\rho{}_{|\mu|\nu]}\mathcal{D}^{[\sigma}L_\sigma{}^{|\mu|\nu]}+4\mathcal{D}_{[\rho}L^\mu{}_{|\nu|\sigma]}\mathcal{D}^{[\rho}L_\mu{}^{|\nu|\sigma]}.
	\end{split}
	\end{align}
	The  above contributions have been explicitly written in terms of disformation rather than non-metricity, so that Eq. \eqref{GBSTEGR} can be easily compared with the corresponding terms in the metric teleparallel case. However, when expanded in terms of non-metricity (as done in Appendix \ref{app:GB}), it turns out that the total number of linearly independent terms used to construct the Gauss-Bonnet invariant is 67, that is  more if compared with the metric teleparallel case, where 56 terms occur. This is expected since the torsion tensor is antisymmetric in its last two indices, which is more restrictive than the symmetries of non-metricity (which instead is symmetric in its two last indices). This is similar to what happens when confronting TEGR (or STEGR) with New General Relativity (NGR) \cite{Hayashi:1979qx} or Newer General Relativity \cite{BeltranJimenez:2017tkd}, where more linearly independent terms consistently appear in the symmetric teleparallel case. This difference is even more extreme when comparing the total EFT-theory of metric and symmetric teleparallel gravity, since the latter consists of 189 terms that are almost the double of the 95 terms appearing in the former. Notice that also in the case of metric teleparallel gravity one might construct some ``symmetric teleparallel Gauss-Bonnet invariant'', which differ from $\mathcal{G}_\mathrm{STEGR}$ by a boundary term. This term would be pseudo diffeomorphism invariant under transformations of the metric alone. Results in \cite{BeltranJimenez:2022azb} suggest that this could be called pseudo-diffeomorphism invariance of type II, which is the symmetric teleparallel analogue to pseudo-Lorentz invariance of type II considered in \cite{Blixt:2022rpl}. Similarly to the case of nonlinear extensions of TEGR, $f(\mathbb{Q})$-theories suffer of the strong coupling problem,  due to accidental symmetry occurring at the linear perturbations around Minkowski \cite{BeltranJimenez:2021auj}. 
	\begin{table}[h]
		\centering
		\begin{tabular}{|c|c|c|c|c|c|c|c|}
			\hline
			$\mathcal{G}_Q$ & EFT $Q^4$ & $\mathcal{G}_{\mathcal{D}Q}$ & EFT $Q^2 \mathcal{D}Q$ & $\mathcal{G}_{\mathcal{DD}Q}$ & EFT $ (\mathcal{D}Q)^2$ & Total $\mathcal{G}_\mathrm{STEGR}$ & Total EFT  \\ \hline
			33 & 69 & 21 & 95 & 13 & 25 & 67  & 189  \\ \hline
		\end{tabular}
		\caption{Number of terms involved in the construction of the symmetric teleparallel Gauss-Bonnet invariant and number of terms in symmetric teleparallel EFT.}
		\label{symtable}
	\end{table}

	\subsection{General teleparallel Gauss-Bonnet invariant}
	Here we consider the general teleparallel case for which the Levi-Civita Riemann tensor can be expressed as
	\begin{align}
	\begin{split}
	\label{eq:TrinityGenGB}
	R^\mu{}_{\nu\rho\sigma}=& -2\mathcal{D}_{[\rho}\tensor{K}{^\mu_|_\nu_|_\sigma_]} - 2 \tensor{K}{^\mu_\lambda_[_\rho}\tensor{K}{^\lambda_|_\nu_|_\sigma_]}  -2\mathcal{D}_{[\rho}\tensor{L}{^\mu_|_\nu_|_\sigma_]} - 2 \tensor{L}{^\mu_\lambda_[_\rho}\tensor{L}{^\lambda_|_\nu_|_\sigma_]}  - 2\tensor{K}{^\mu_\lambda_[_\rho}\tensor{L}{^\lambda_|_\nu_|_\sigma_]} \\
	&- 2\tensor{L}{^\mu_\lambda_[_\rho}\tensor{K}{^\lambda_|_\nu_|_\sigma_]}\\
	&=\overset{T}{R}^\mu{}_{\nu\rho\sigma}+\overset{Q}{R}^\mu{}_{\nu\rho\sigma}- 2K^\mu{}_{\lambda[\rho}L^\lambda{}_{|\nu|\sigma]} - 2L^\mu{}_{\lambda[\rho}K^\lambda{}_{|\nu|\sigma]}.
	\end{split}
	\end{align}
	Thus,
	\begin{align}
	\begin{split}
	R^\mu{}_{\nu\rho\sigma}=\overset{T}{R}^\mu{}_{\nu\rho\sigma}+\overset{Q}{R}^\mu{}_{\nu\rho\sigma}- 2K^\mu{}_{\lambda[\rho}L^\lambda{}_{|\nu|\sigma]} - 2L^\mu{}_{\lambda[\rho}K^\lambda{}_{|\nu|\sigma]}.
	\end{split}
	\end{align}
	In the general case of teleparallelism, the Gauss-Bonnet invariant can be written as\footnote{Note that without imposing torsion or nonmetricity equals zero $\mathcal{G}_{\mathrm{TEGR}}\neq \mathcal{G}\neq \mathcal{G}_{\mathrm{STEGR}}$.} 
	\begin{align}
	\begin{split}
	\mathcal{G}&=\mathcal{G}_{\mathrm{GTEGR}}=\mathcal{G}_{\mathrm{TEGR}}+\mathcal{G}_{\mathrm{STEGR}}+\mathcal{G}_{TTTQ}+\mathcal{G}_{TTQQ}+\mathcal{G}_{TQQQ}\\
	&+\mathcal{G}_{TQ\mathcal{D}T}+\mathcal{G}_{QQ\mathcal{D}T}+\mathcal{G}_{TT\mathcal{D}Q}+\mathcal{G}_{TQ\mathcal{D}Q}+\mathcal{G}_{\mathcal{D}T\mathcal{D}Q},
	\end{split}
	\end{align}
	with
	\begin{align}
	\begin{split}
	\mathcal{G}_{TTTQ}&:=8K^\lambda{}_{\kappa[\lambda}K^\kappa{}_{|\mu|\nu]}K^\alpha{}_{\beta[\alpha}L^\beta{}_{|\rho|\sigma]}g^{\mu\nu}g^{\rho\sigma}+8K^\lambda{}_{\kappa[\lambda}K^\kappa{}_{|\mu|\nu]}L^\alpha{}_{\beta[\alpha}K^\beta{}_{|\rho|\sigma]}g^{\mu\nu}g^{\rho\sigma}\\
	&-32K^\lambda{}_{\kappa[\lambda}K^\kappa{}_{|\mu|\nu]}K_\alpha{}^{\beta[\alpha}L_\beta{}^{|\rho|\sigma]}-32K^\lambda{}_{\kappa[\lambda}K^\kappa{}_{|\mu|\nu]}L_\alpha{}^{\beta[\alpha}K_\beta{}^{|\rho|\sigma]}\\
	&+8K_{\mu\kappa[\rho}K^\kappa{}_{|\nu|\sigma]}K^{\mu\lambda[\rho}L_\lambda{}^{|\nu|\sigma]}+8K_{\mu\kappa[\rho}K^\kappa{}_{|\nu|\sigma]}L^{\mu\lambda[\rho}K_\lambda{}^{|\nu|\sigma]},
	\end{split}
	\end{align}
	\begin{align}
	\begin{split}
	\mathcal{G}_{TTQQ}&:=4K^\kappa{}_{\lambda[\kappa}L^\lambda{}_{|\mu|\nu]}K^\alpha{}_{\beta[\alpha}L^\beta{}_{|\rho|\sigma]}g^{\mu\nu}g^{\rho\sigma}+8K^\kappa{}_{\lambda[\kappa}L^\lambda{}_{|\mu|\nu]}L^\alpha{}_{\beta[\alpha}K^\beta{}_{|\rho|\sigma]}g^{\mu\nu}g^{\rho\sigma}\\
	&+4L^\kappa{}_{\lambda[\kappa}K^\lambda{}_{|\mu|\nu]}L^\alpha{}_{\beta[\alpha}K^\beta{}_{|\rho|\sigma]}g^{\mu\nu}g^{\rho\sigma}-16K^\kappa{}_{\lambda[\kappa}L^\lambda{}_{|\mu|\nu]}K_\alpha{}^{\beta[\alpha}L_\beta{}^{|\rho|\sigma]}\\
	&-32K^\kappa{}_{\lambda[\kappa}L^\lambda{}_{|\mu|\nu]}L_\alpha{}^{\beta[\alpha}K_\beta{}^{|\rho|\sigma]}-16L^\kappa{}_{\lambda[\kappa}K^\lambda{}_{|\mu|\nu]}L_\alpha{}^{\beta[\alpha}K_\beta{}^{|\rho|\sigma]}\\
	&+4K_{\mu\kappa[\rho}L^\kappa{}_{|\nu|\sigma]}K^{\mu\lambda[\rho}L_\lambda{}^{|\nu|\sigma]}+8K_{\mu\kappa[\rho}L^\kappa{}_{|\nu|\sigma]}L^{\mu\lambda[\rho}K_\lambda{}^{|\nu|\sigma]}\\
	&+4L_{\mu\kappa[\rho}K^\kappa{}_{|\nu|\sigma]}L^{\mu\lambda[\rho}K_\lambda{}^{|\nu|\sigma]},
	\end{split}
	\end{align}
	\begin{align}
	\begin{split}
	\mathcal{G}_{TQQQ}&:=8L^\lambda{}_{\kappa[\lambda}L^\kappa{}_{|\mu|\nu]}L^\alpha{}_{\beta[\alpha}K^\beta{}_{|\rho|\sigma]}g^{\mu\nu}g^{\rho\sigma}+8L^\lambda{}_{\kappa[\lambda}L^\kappa{}_{|\mu|\nu]}K^\alpha{}_{\beta[\alpha}L^\beta{}_{|\rho|\sigma]}g^{\mu\nu}g^{\rho\sigma}\\
	&-32L^\lambda{}_{\kappa[\lambda}L^\kappa{}_{|\mu|\nu]}L_\alpha{}^{\beta[\alpha}K_\beta{}^{|\rho|\sigma]}-32L^\lambda{}_{\kappa[\lambda}L^\kappa{}_{|\mu|\nu]}K_\alpha{}^{\beta[\alpha}L_\beta{}^{|\rho|\sigma]}\\
	&+8L_{\mu\kappa[\rho}L^\kappa{}_{|\nu|\sigma]}L^{\mu\lambda[\rho}K_\lambda{}^{|\nu|\sigma]}+8L_{\mu\kappa[\rho}L^\kappa{}_{|\nu|\sigma]}K^{\mu\lambda[\rho}L_\lambda{}^{|\nu|\sigma]},
	\end{split}
	\end{align}
	\begin{align}
	\begin{split}
	\mathcal{G}_{TQ\mathcal{D}T}&:=8\mathcal{D}_{[\kappa}K^\kappa{}_{|\mu|\nu]}K^\alpha{}_{\lambda[\alpha}L^\lambda{}_{|\rho|\sigma]}g^{\mu\nu}g^{\rho\sigma}+8\mathcal{D}_{[\kappa}K^\kappa{}_{|\mu|\nu]}L^\alpha{}_{\lambda[\alpha}K^\lambda{}_{|\rho|\sigma]}g^{\mu\nu}g^{\rho\sigma}\\
	&-32\mathcal{D}_{[\rho}K^\rho{}_{|\mu|\nu]}K_\sigma{}^{\lambda[\sigma}L_\lambda{}^{|\mu|\nu]}-32\mathcal{D}_{[\rho}K^\rho{}_{|\mu|\nu]}L_\sigma{}^{\lambda[\sigma}K_\lambda{}^{|\mu|\nu]}\\
	&+8\mathcal{D}_{[\rho}K_{|\mu\nu|\sigma}K^{\mu\lambda[\rho}L_\lambda{}^{|\nu|\sigma]}+8\mathcal{D}_{[\rho}K_{|\mu\nu|\sigma}L^{\mu\lambda[\rho}K_\lambda{}^{|\nu|\sigma]},
	\end{split}
	\end{align}
	\begin{align}
	\begin{split}
	\mathcal{G}_{QQ\mathcal{D}T}&:=8\mathcal{D}_{[\kappa}K^\kappa{}_{|\mu|\nu]}L^\alpha{}_{\lambda[\alpha}L^\lambda{}_{|\rho|\sigma]}g^{\mu\nu}g^{\rho\sigma}-32\mathcal{D}_{[\rho}K^\rho{}_{|\mu|\nu]}L_\sigma{}^{\lambda[\sigma}L_\lambda{}^{|\mu|\nu]}\\
	&+8\mathcal{D}_{[\rho}K_{|\mu\nu|\sigma}L^{\mu\lambda[\rho}L_\lambda{}^{|\nu|\sigma]},
	\end{split}
	\end{align}
	\begin{align}
	\begin{split}
	\mathcal{G}_{TT\mathcal{D}Q}&:=8\mathcal{D}_{[\kappa}L^\kappa{}_{|\mu|\nu]}K^\alpha{}_{\lambda[\alpha}K^\lambda{}_{|\rho|\sigma]}g^{\mu\nu}g^{\rho\sigma}-32\mathcal{D}_{[\rho}L^\rho{}_{|\mu|\nu]}K_\sigma{}^{\lambda[\sigma}K_\lambda{}^{|\mu|\nu]}\\
	&+8\mathcal{D}_{[\rho}L_{|\mu\nu|\sigma}K^{\mu\lambda[\rho}K_\lambda{}^{|\nu|\sigma]},
	\end{split}
	\end{align}
	\begin{align}
	\begin{split}
	\mathcal{G}_{TQ\mathcal{D}Q}&:=8\mathcal{D}_{[\kappa}L^\kappa{}_{|\mu|\nu]}K^\alpha{}_{\lambda[\alpha}L^\lambda{}_{|\rho|\sigma]}g^{\mu\nu}g^{\rho\sigma}+8\mathcal{D}_{[\kappa}L^\kappa{}_{|\mu|\nu]}L^\alpha{}_{\lambda[\alpha}K^\lambda{}_{|\rho|\sigma]}g^{\mu\nu}g^{\rho\sigma}\\
	&-32\mathcal{D}_{[\rho}L^\rho{}_{|\mu|\nu]}K_\sigma{}^{\lambda[\sigma}L_\lambda{}^{|\mu|\nu]}-32\mathcal{D}_{[\rho}L^\rho{}_{|\mu|\nu]}L_\sigma{}^{\lambda[\sigma}K_\lambda{}^{|\mu|\nu]}\\
	&+8\mathcal{D}_{[\rho}L_{|\mu\nu|\sigma}K^{\mu\lambda[\rho}L_\lambda{}^{|\nu|\sigma]}+8\mathcal{D}_{[\rho}L_{|\mu\nu|\sigma}L^{\mu\lambda[\rho}K_\lambda{}^{|\nu|\sigma]},
	\end{split}
	\end{align}
	\begin{align}
	\begin{split}
	\mathcal{G}_{\mathcal{D}T\mathcal{D}Q}&:=4\mathcal{D}_{[\kappa}K^\kappa{}_{|\mu|\nu]}\mathcal{D}_{[\lambda}L^\lambda{}_{|\rho|\sigma]}g^{\mu\nu}g^{\rho\sigma}-16\mathcal{D}_{[\rho}K^\rho{}_{|\mu|\nu]}\mathcal{D}^{[\sigma}L_\sigma{}^{|\mu|\nu]}\\
	&+4\mathcal{D}_{[\rho}K_{|\mu\nu|\sigma]}\mathcal{D}^{[\rho}L^{|\mu\nu|\sigma]}.
	\end{split}
	\end{align}
	Note that although $\mathcal{G}_\mathrm{TEGR},\mathcal{G}_\mathrm{STEGR},$ and $\mathcal{G}_\mathrm{GTEGR}$ all equals the Gauss-Bonnet invariant $\mathcal{G}$ under their respective geometrical assumptions, we still have $\mathcal{G}_\mathrm{TEGR}\neq\mathcal{G}_\mathrm{STEGR}\neq\mathcal{G}_\mathrm{GTEGR}$. From the above computations, it turns out that the number of possible invariant terms that can be constructed in the general teleparallel geometry and with the Levi-Civita covariant derivative is 1052 terms. Also notice that such terms can be properly rewritten by performing integration by parts, which gives rise to boundary terms. The non-uniqueness of the boundary term used in \cite{Kofinas:2014owa} was already addressed in \cite{Gonzalez:2015sha}. Furthermore, we calculated the number of invariant terms grouped by the order of Levi-Civita covariant derivative, torsion and non-metricity. 
	\begin{table}
		\centering
		\noindent
		\begin{tabular}{|c|c|c|c|c|c|c|c|c|c|c|c|c|c|}
			\hline
			\multicolumn{14}{|c|}{EFT}  \\ \hline
			$T^4$  &  $Q^4$ &  $T^2 \mathcal{D}T$ &  $Q^2 \mathcal{D}Q$ &  $ (\mathcal{D}T)^2$   & $ (\mathcal{D}Q)^2$   & $T^3Q$  & $T^2Q^2$ &  $TQ^3 $ &  $TQ \mathcal{D}T$ &  $ Q^2 \mathcal{D}T$ &  $ TQ \mathcal{D}Q$  &   $ T^2 \mathcal{D}Q$  &  $ \mathcal{D}T \mathcal{D}Q$   \\ \hline
			33  & 69 & 46 & 95 & 16 & 25  & 91  & 184  & 127 & 102 & 61 & 123 & 56 & 24 \\ \hline
			\multicolumn{14}{|c|}{In total 1052 terms}  \\ \hline
		\end{tabular}
		\caption{Number of terms in general teleparallel EFT.}
		\label{genEFTtable}
	\end{table}
	\newline {} \\ It turns out that about half of the independent terms in the general teleparallel case are needed for the construction of the Gauss-Bonnet invariant. This particular combination, which reproduces quantities that can be expressed as the Levi-Civita Riemann tensor and its contractions, is both invariant under diffeomorphisms and Lorentz transformations of both type I and type II (following the definitions in \cite{Blixt:2022rpl}). By changing the coefficients or adding invariant terms that do not take a role, when assembling the Gauss-Bonnet invariant, it is possible to break the symmetries of GR and extract new degrees of freedom in the theory. This needs to be done with caution, since it may lead to the appearance of ghostly degrees of freedom or strongly coupled fields \cite{BeltranJimenez:2021auj}. However, the large number of independent terms may also provide more freedom not only in terms of breaking symmetries but also in the opposite direction (that is greater freedom in constructing a theory with a desired remnant symmetry).    
	\begin{table}
		\centering
		\begin{tabular}{|c|c|c|c|c|c|c|c|c|c|c|c|c|c|}
			\hline
			$\mathcal{G}_T$ & $\mathcal{G}_Q$ &  $\mathcal{G}_{TTTQ}$ & $\mathcal{G}_{TTQQ}$ & $\mathcal{G}_{TQQQ}$ & $\mathcal{G}_{\mathcal{D}T}$ & $\mathcal{G}_{\mathcal{D}Q}$ &   $\mathcal{G}_{TQ\mathcal{D}T}$ & $ \mathcal{G}_{QQ\mathcal{D}T}$ &  $\mathcal{G}_{TQ\mathcal{D}Q}$ & $\mathcal{G}_{TT\mathcal{D}Q}$ & $\mathcal{G}_{\mathcal{DD}T}$ & $\mathcal{G}_{\mathcal{DD}Q}$ &    $\mathcal{G}_{\mathcal{D}T\mathcal{D}Q}$  \\ \hline
			27 & 33 & 70 & 100 & 65 & 18 & 21 & 30 & 27 & 42 &  33 & 11 & 13 & 16  \\ \hline
			\multicolumn{14}{|c|}{In total 506 terms}  \\ \hline
		\end{tabular}
		\caption{Number of terms involved in the construction of the general teleparallel Gauss-Bonnet invariant.}
		\label{genGBtable}
	\end{table}

	\section{Constructing teleparallel theories from fundamental principles}
	\label{sec:Principles}
	It is clear from Table \ref{teletable},\ref{symtable},\ref{genEFTtable}, and \ref{genGBtable} that it is easy to construct modified theories of gravity within the teleparallel set up. Yet extensions including additional fields, higher dimensions, non-minimal couplings to matter, or parity violation were not considered in this article. In four dimensions, the Gauss-Bonnet invariant is purely topological and this fact can be used to reduce the number of independent terms by one, which is a very tiny achievement. Considering even higher order invariants will obviously be even worse and, in this context, theories like $f(\mathbb{T})$ or $f(\mathbb{T},T_\mathcal{G})$-gravity \cite{Kofinas:2014owa} will appear to be very \emph{ad-hoc} without a clear fundamental guiding principle. The importance of guiding principles in modified gravity has also been stressed in \cite{Golovnev:2022bfm,Shankaranarayanan:2022wbx}. Nonetheless, such theories may still turn out to be good models for describing gravity, but their viability must be tested by experiments and/or observations.
	
	Usually, in teleparallel-like models, one typically either encounters ghost-instabilities \cite{BeltranJimenez:2019odq,Ortin:2015hya,Li:2022mti} or strongly coupled fields \cite{BeltranJimenez:2019nns,BeltranJimenez:2020fvy,BeltranJimenez:2019tme,BeltranJimenez:2021auj} when going beyond the  geometric  trinity of gravity. Theories that are plagued by ghost-instabilities are typically excluded by observational predictions, since ghost fields are not expected in nature. On the other hand, the presence of strongly coupled fields is more difficult to notice, but it is at least known for the cases of $f(\mathbb{T})$-gravity \cite{BeltranJimenez:2020fvy, Golovnev:2020nln, Bahamonde:2022ohm}, $f(\mathbb{Q})$-gravity \cite{BeltranJimenez:2019tme} and NGR \cite{BeltranJimenez:2019nns}. Moreover, many observational predictions have been carried out for those theories on backgrounds with strongly coupled fields \cite{Bahamonde:2021gfp}. However, these predictions are not reliable due to the strong coupling problem, since theories with strongly coupled fields might even suffer from ghost-instabilities \cite{Annala:2022gtl}, which is not manifested in highly symmetric backgrounds. From the Hamiltonian perspective, the presence of strongly coupled fields seems to be connected with the presence of bifurcations in the classification of constraints \cite{Cheng:1988zg,Li:2011rn,Blagojevic:2020dyq,Ferraro:2020tqk,Chakrabortty:2022giq,Chen:1998ad}. It is clear that this only appears when the symmetries of GR are \textit{partially} broken, since in the other cases all constraints are of first class with respect to any background (see \cite{Mitric:2019rop} for the examples on generic NGR where Lorentz invariance of type II is fully broken, or Refs. \cite{Blagojevic:2000qs, DAmbrosio:2020nqu} respectively for TEGR and STEGR, where the symmetries of GR are preserved up to a boundary term). The cases in which the symmetries of GR are partially broken result in some remnant symmetries \cite{Ferraro:2020tqk} that are typically not easy to grasp. This is not a great surprise from the group theory point of view. As a matter of fact, taking the Lorentz group as an example, it is known that it can be generated by Lorentz boosts and Lorentz rotations. However, the set of Lorentz boosts does not form a group itself, while the set of Lorentz rotations does. Thus, it is clear that the remnant symmetry of the modified teleparallel theory does not necessarily form a group which can give rise to rather peculiar behaviors in the Hamiltonian analysis, like bifurcations in the classification of constraints. Hence, a guiding principle in the construction of modified theories of gravity with remnant symmetries could be requiring that the symmetry forms a group, like Lorentz rotations or 3D diffeomorphisms (see Ref.  \cite{Mukohyama:2019unx} for the application of this principle). In the case of teleparallel gravity, one may investigate if higher order terms can be used to form a well-behaved remnant symmetry.
	
	The EH action formulation includes divergences that have led to the introduction of the York-Gibbons-Hawking boundary term \cite{York:1972sj,Gibbons:1976ue}, which aims to remove the divergences. When the action is supplemented with this boundary term, the (new) theory predicts an ADM-mass. In the teleparallel part of trinity of gravity, the procedure is different as, in these formulations, many redundant fields are included and the calculation of the mass is gauge dependent \cite{Gomes:2022vrc}. The York-Gibbons-Hawking boundary term can be replaced by a gauge choice and is therefore no longer necessary \cite{Oshita:2017nhn}. For this reason, the principle of always choosing the so-called \textit{canonical frame} was introduced \cite{BeltranJimenez:2019bnx}. It is defined as the frame for which the energy-momentum tensor associated with the metric vanishes, but consequences that this principle gives in modified theories of gravity has yet to be investigated.

	\section{Discussion and Conclusions}
	\label{sec:Conclusions}
	
	In this paper,  the explicit expression of the teleparallel equivalent for the Gauss-Bonnet invariant has been explicitly given for the trinity of gravity (metric teleparallel, symmetric teleparallel, and general teleparallel equivalence). In the standard Riemannian formulation of quadratic gravity, the Gauss-Bonnet invariant only consists of three terms. In four dimensions, the Gauss-Bonnet invariant is a topological identity that can be used to restricts the number of possible constructions in quadratic gravity.
	
	We find that there are far more terms in teleparallel quadratic gravity theories and that, from an EFT perspective, one may construct 95 terms in metric teleparallel gravity, 189 terms in symmetric teleparallel gravity, and 1052 terms in general teleparallel gravity. The Gauss-Bonnet invariant is thus not-trivial and involves 56 terms in metric teleparallel gravity, 67 terms in symmetric teleparallel gravity and 506 terms in general teleparallel gravity. The results are summarized in Table \ref{teletable},\ref{symtable},\ref{genEFTtable}, and \ref{genGBtable} and the detailed expressions are reported  in Appendix \ref{app:GB}.
	
	Theoretical guiding principles (which are relevant in particular in the field of teleparallel gravity) was discussed in section \ref{sec:Principles}. These principles are additional to the typical demands such as no ghosts or strongly coupled fields, consistency with observations, Occam's razor, \emph{etc.} and can be summarized as follows
	\begin{enumerate}
		\item Modified teleparallel theories should be constructed from partially (and not fully) breaking some symmetries of GR.
		\item Partially breaking a symmetry leads to a remnant symmetry, required to be well-behaved.
		\item Other principles like predicting the ADM-mass from the gauge fixing, coinciding with the \textit{canonical frame}, are worth exploring. 
	\end{enumerate}
	The first principle is a consequence of the fact that ghosts generically appear in teleparallel theories of gravity \cite{BeltranJimenez:2019odq,Ortin:2015hya,Li:2022mti}. The second principle is related to the appearance of strongly coupled fields for which the Dirac-Bergmann algorithm becomes background-dependent when classifying the constraints into first and second class \cite{Cheng:1988zg,Li:2011rn,Blagojevic:2020dyq,Ferraro:2020tqk,Chakrabortty:2022giq,Chen:1998ad}. Perhaps this could be avoided by demanding that the generators of the remnant symmetry form a group (note that one still needs to check for ghosts in this case). The third principle gives the prescription to obtain a well-defined mass in gravity, in a way that is less \emph{ad-hoc} than the conventional procedure by York, Gibbons, and Hawking. These are principles that we find worth exploring after reflecting on past results, which can hopefully provide a criterion to focus on viable models within the plethora of modified gravity theories to investigate further. 
	
	In this perspective, $f(\mathbb{T})$ is poorly motivated and $f(\mathbb{T},T_{\mathcal{G}})$ even worse. Among higher-order torsion theories, they are very specific and break the principles listed above, except for the first one (thus seemingly avoiding ghosts) at first sight. However, regarding the second principle, $f(\mathbb{T})$-gravity evidently suffers from strong coupling \cite{BeltranJimenez:2020fvy,Golovnev:2020nln,Bahamonde:2022ohm} and there are no indications suggesting that the same would not occur in $f(\mathbb{T},T_\mathcal{G})$. Although there are many ways to construct higher order torsion theories like $f(\mathbb{T})$ and $f(\mathbb{T},T_\mathcal{G})$, the selection of these particular models lacks a clear motivation\footnote{However, it should be stressed that, in principle, they could still turn out to be viable theories. Nevertheless, to gain a more profound understanding, it is necessary to investigate how the strongly coupled fields affect the theory.}. This is particurlarly evident in $f(\mathbb{T},T_\mathcal{G})$ gravity, where $T_\mathcal{G}$ differs from the Gauss-Bonnet invariant by a boundary term that is clearly not unique. On the other hand, the community has gained significant insights into $f(\mathbb{T})$-gravity and the results provided in this paper might serve as an interesting example to further explore the strong coupling problem.
	
	\section*{Acknowledgments}
This paper is based upon work from COST Action CA21136 {\it Addressing observational tensions in cosmology with systematics and fundamental physics} (CosmoVerse) supported by European Cooperation in Science and Technology.
	The authors acknowledge the Istituto Nazionale di Fisica Nucleare (INFN), Sezione di Napoli, \textit{iniziative specifiche} GINGER,  QGSKY, and MOONLIGHT2. D.B. is grateful for the very interesting discussions that were held at the Metric-Affine Gravity conference in Tartu, 2022.

	\appendix

	\section{Gauss-Bonnet terms explicitly in torsion and non-metricity}
	\label{app:GB}
	The Gauss-Bonnet terms of metric and symmetric teleparallel gravity take part in the construction o general teleparallel theories. Let us now evaluate all the quantities involved in the definition of $\mathcal{G}$, as presented in Eq. \eqref{eq:TeleGB}. Starting with the metric teleparallel case, the term $\mathcal{G}_T$ explicitly reads
	\begin{align}
	\begin{split}
	\mathcal{G}_T&=\frac{1}{4}T_{\alpha\lambda\mu}T_{\gamma\nu\rho}T^{\lambda\alpha\mu}T^{\nu\gamma\rho}+\frac{1}{16}T_{\alpha\lambda\mu}T^{\alpha\lambda\mu}T^{\gamma\nu\rho}T_{\gamma\nu\rho}+\frac{1}{4}T_{\alpha\lambda\mu}T^{\alpha\lambda\mu}T_{\gamma\nu\rho}T^{\nu\gamma\rho}\\
	&-T_{\alpha\lambda\mu}T^{\alpha\nu\rho}T^{\lambda\gamma\mu}T_{\nu\gamma\rho}-\frac{1}{8}T_{\alpha\lambda\mu}T^{\alpha\nu\rho}T^{\gamma\lambda\mu}T_{\gamma\nu\rho}-T_{\alpha\lambda\mu}T^{\alpha\nu\rho}T^{\gamma\lambda\mu}T_{\nu\gamma\rho}\\
	&+2T_{\alpha\lambda\mu}T^{\lambda\gamma\mu}T^{\nu\alpha\gamma}T_\nu-\frac{1}{2}T_\lambda T_{\gamma\mu\nu}T^{\gamma\mu\nu}T^\lambda-T_\lambda T_{\gamma\mu\nu}T^{\mu\gamma\nu}T^\lambda+T_\lambda T^\mu T_\lambda T^\mu\\
	&-T_{\lambda\alpha\gamma}T^{\mu\alpha\gamma}T^\lambda T_\mu+2T_{\alpha\gamma\lambda}T^{\gamma\mu\nu}T_{\mu}{}^{\alpha}{}_\nu T_\lambda+4T_{\alpha}{}^\gamma{}_{\lambda}T^{\alpha\mu\nu}T_{(\gamma\mu)\nu}T^\lambda-4T_{\alpha\gamma\lambda}T^{(\alpha\gamma)\mu}T^\lambda T_\mu\\
	&-\frac{3}{8}T_{\sigma\mu\nu}T^{\sigma\alpha\gamma}T_{\tau\gamma}{}^\nu T^{\tau}{}_\alpha{}^\mu-\frac{3}{4}T_{\alpha\gamma\sigma}T^{\gamma\mu\tau}T_{\mu\nu}{}^{\sigma}T^{\nu\alpha}{}_\tau+\frac{1}{2}T_{\alpha\gamma\sigma}T^{\gamma\alpha\tau}T^{\mu\nu\sigma}T_{\nu\mu\tau}\\
	&-T_{\alpha\gamma\sigma}T^{\gamma\mu\tau}T^\sigma{}_{\mu\nu}T_\tau{}^{\alpha\nu}-\frac{3}{2}T_{\alpha\gamma\sigma}T^{\alpha\mu}{}_\tau T^{\gamma\nu\tau}T_{\mu\nu}{}^\sigma-T_{\alpha\gamma\sigma}T^{\alpha\mu}{}_\tau T^{\gamma\nu\tau}T_{\nu\mu}{}^\sigma\\
	&+\frac{1}{2}T_{\alpha\gamma\sigma}T^{\gamma\alpha\tau}T^{\sigma\mu\nu}T_{\tau\mu\nu}+\frac{1}{2}T_{\alpha\gamma\sigma}T^{\alpha\gamma\tau}T^{\mu\nu\sigma}T_{\mu\nu\tau}\\
	&+T_{\alpha\gamma\sigma}T^{\alpha\gamma\tau}T^{\mu\nu\sigma}T_{\nu\mu\tau}+\frac{1}{2}T_{\alpha\gamma\sigma}T^{\alpha\gamma\tau}T^{\sigma\mu\nu}T_{\tau\mu\nu}+\frac{1}{2}T_{\alpha\gamma\sigma}T^{\mu\nu\sigma}T^{\tau\gamma}{}_\nu T_\tau{}^\alpha{}_\mu,
	\end{split}
	\end{align}
	consisting of 27 terms. The $\mathcal{G}_{\mathcal{D}T}$ term explicitly reads 
	\begin{align}
	\begin{split}
	\mathcal{G}_{\mathcal{D}T}&=-2\mathcal{D}_\alpha T_{\beta\kappa\lambda}T^{(\kappa\alpha)\mu}T^{\lambda\beta}{}_\mu-\mathcal{D}_\alpha T_{\beta\kappa\lambda}T^{\alpha\kappa\mu}T^{\beta\lambda}{}_\mu-\frac{1}{2}\mathcal{D}_\alpha T^{\lambda}T^{\alpha\mu\nu}T_{\lambda\mu\nu}\\
	&-\mathcal{D}_\alpha T^{\beta\alpha\lambda}T_{\beta\mu\nu}T_\lambda{}^{\mu\nu}+2\mathcal{D}_\alpha T_{\beta\kappa\lambda}T^{\beta\kappa}{}_\mu T^{[\mu\lambda]\alpha}-\mathcal{D}_\alpha T_{\beta\kappa\lambda}T^{\kappa\beta\mu}T_\mu{}^{\alpha\lambda}\\
	&+2\mathcal{D}_\alpha T^{\lambda}T_{\lambda\mu\nu}T^{\mu\alpha\nu}-\mathcal{D}_\alpha T^{\beta\alpha\lambda}T_{\lambda}{}^{\mu\nu}T_{\mu\beta\nu}+\frac{1}{2}\mathcal{D}_\alpha T_{\beta\kappa\lambda}T^{\alpha\beta\mu}T_\mu{}^{\kappa\lambda}-\mathcal{D}_\alpha T^{\lambda}T^{\alpha\mu\nu}T_{\mu\lambda\nu}\\
	&-\mathcal{D}_\alpha T^{\beta\alpha\lambda}T_{\beta}{}^{\mu\nu}T_{\mu\lambda\nu}+\mathcal{D}_\alpha T^{\alpha\kappa\lambda}T_{\kappa}{}^{\mu\nu}T_{\mu\lambda\nu}+4\mathcal{D}_\alpha T^{\beta\alpha\lambda}T_{(\beta\lambda)\mu}T^\mu+4\mathcal{D}_\alpha T_{\lambda}T^{[\alpha\lambda]\mu}T_\mu,
	\end{split}
	\end{align}
	with 18 terms. The $\mathcal{G}_{\mathcal{DD}T}$ term explicitly reads
	\begin{align}
	\begin{split}
	\mathcal{G}_{\mathcal{DD}T}&=\mathcal{D}_\mu T_{\rho\nu\sigma}\mathcal{D}^\mu T^{\nu\rho\sigma}+\frac{3}{2}\mathcal{D}_\mu T_{\nu\rho\sigma}\mathcal{D}^\mu T^{\nu\rho\sigma}-2\mathcal{D}_\nu T_{\rho\mu\sigma}\mathcal{D}^\mu T^{\nu\rho\sigma}-\frac{3}{2}\mathcal{D}_\nu T^\mu{}_{\rho\sigma}\mathcal{D}_\mu T^{\nu\rho\sigma}\\
	&+3\mathcal{D}_\rho T_{\nu\sigma}{}^\mu \mathcal{D}_\mu T^{\nu\rho\sigma}+\mathcal{D}_\rho T_{\sigma\nu}{}^\mu \mathcal{D}_\mu T^{\nu\rho\sigma}+4\mathcal{D}_\nu T^{\rho\nu\mu}\mathcal{D}_\mu T_\rho\\
	&-4\mathcal{D}_\mu T_{\rho}\mathcal{D}^\mu T^\rho+8\mathcal{D}_\nu T^{(\mu\nu)\rho}\mathcal{D}_\mu T_\rho+4\mathcal{D}_\mu T^\mu\mathcal{D}_\nu T^\nu.
	\end{split}
	\end{align}
	with 11 terms.
	
	In the case of symmetric teleparallel gravity, the Gauss-Bonnet invariant consists of the sum of three quantities defined in Eq. \eqref{GBSTEGR}. They explicitly read
	\begin{align}
	\begin{split}
	\mathcal{G}_Q&=-\frac{1}{4}\Bar{Q}_\mu Q^\mu Q_\nu Q^\nu+\frac{1}{4}\Bar{Q}_\mu Q^\mu Q^{\nu\sigma\rho}Q_{\nu\sigma\rho}-\frac{1}{2}\Bar{Q}_\alpha Q^{\alpha}Q^{\nu\sigma\rho}Q_{\sigma\nu\rho}+\Bar{Q}_\mu Q^\mu \Bar{Q}_\nu Q^\nu \\
	&+\frac{1}{16}Q_\mu Q^\mu Q_\nu Q^\nu+\frac{1}{8}Q_\mu Q^\mu Q^{\sigma\nu\rho}Q_{\sigma\nu\rho}-\frac{1}{4}Q_\mu Q^\mu  Q^{\sigma\nu\rho}Q_{\nu\sigma\rho}+Q_\mu  Q^{\sigma\mu\kappa}Q_{\sigma\nu\rho}Q^{\kappa\nu\rho}\\
	&-Q_\mu  Q^{\sigma\mu\kappa}Q_{(\sigma\kappa)\nu}Q^{\nu}+8Q_\mu  Q^{\sigma\mu\kappa}Q_{\nu\sigma\rho}Q^{[\rho\nu]}{}_\kappa-\frac{1}{2}Q_\mu  Q^{\mu\sigma\kappa}Q_{\sigma\nu\rho}Q_\kappa{}^{\nu\rho}\\
	&+2Q_\mu  Q^{\mu\sigma\kappa}Q_{\sigma\kappa\nu}Q^{\nu}-\frac{1}{2}Q_\mu  Q^{\mu\sigma\kappa}Q_{\nu\sigma\kappa}Q^\nu+4Q_\mu  Q^{\mu\sigma}{}_\kappa Q_{\nu\sigma\rho}Q^{[\nu\rho]\kappa}\\
	&-  Q^{\mu\gamma\sigma}Q_{\gamma\mu\kappa}Q_{\sigma\nu\rho}Q^{\kappa\nu\rho}+\frac{1}{16}Q^{\mu\gamma\sigma}Q_{\mu\gamma\sigma}Q^{\kappa\nu\rho}Q_{\kappa\nu\rho}+Q^{\mu\gamma\sigma}Q_{\gamma\mu\kappa}Q^{\sigma\nu\rho}Q_{\nu\kappa\rho}\\
	&+\frac{1}{2}Q^{\mu\gamma\sigma}Q_{[\gamma\mu]\sigma}Q^{\kappa\nu\rho}Q_{\nu\kappa\rho}-\frac{1}{4}Q_\mu{}^{\gamma\sigma}Q_{\gamma\kappa\nu}Q^{\kappa}{}_{\sigma\rho}Q^{\rho\mu\nu}+\frac{3}{2}Q^{\mu\gamma\sigma}Q_{\mu\gamma\kappa}Q_\sigma{}^{\mu\rho}Q^\kappa{}_{\nu\rho}\\
	&-Q^{\mu\gamma\sigma}Q_{\mu\gamma\kappa}Q_\sigma{}^{\nu\rho}Q_{\nu\rho}{}^\kappa-\frac{1}{2}Q^{\mu\gamma}{}_\sigma Q_{\mu\gamma\kappa}Q^{\nu\sigma\rho}Q_{\nu\rho}{}^\kappa+2Q^{\mu\gamma}{}_\sigma Q_{\mu\gamma\kappa}Q^{\nu\sigma\rho}Q_{\rho\nu}{}^\kappa\\
	&-\frac{3}{2}Q^{\mu\gamma\sigma}Q_{\gamma\mu\kappa}Q_{\nu\sigma\rho}Q^{\rho\kappa\nu}-\frac{1}{2}Q^{\mu\gamma\sigma}Q_{\mu\kappa\nu}Q_{\gamma}{}^{\kappa\rho}Q_{\sigma\rho}{}^\nu+Q^{\mu\gamma\sigma}Q_{\mu\kappa\nu}Q_\gamma{}^{\kappa\rho}Q_{\rho\sigma}{}^\nu\\
	&-\frac{3}{8}Q^{\mu\gamma\sigma}Q_{\mu\kappa\nu}Q_{\rho\gamma}{}^\kappa Q^{\rho\nu}{}_\sigma-\frac{1}{2}Q^{\mu\gamma\sigma}Q_{\mu\kappa\nu}Q_{\gamma\sigma\rho}Q^{\rho\kappa\nu}-\frac{1}{8}Q^{\mu\gamma\sigma}Q_{\mu\kappa\nu}Q_{\rho\gamma\sigma}Q^{\rho\kappa\nu},
	\end{split}
	\end{align}
	with 33 terms. The $\mathcal{G}_{\mathcal{D}Q}$ invariant explicitly reads
	\begin{align}
	\begin{split}
	\mathcal{G}_{\mathcal{D}Q}&=2\mathcal{D}_\alpha Q_{\beta\gamma\delta}Q^{\alpha\gamma\kappa}Q^{\beta\delta}{}_\kappa+2\mathcal{D}_\alpha Q_{\beta\gamma\delta}Q^{\gamma\alpha\kappa}Q^{\delta\beta}{}_\kappa-\frac{13}{2}\mathcal{D}_\alpha Q_{\beta\gamma\delta} Q^{\beta\alpha\kappa}Q^{\gamma\delta}{}_\kappa\\
	&-\mathcal{D}_\alpha \Bar{Q}^\delta Q^{\alpha\epsilon\kappa}Q_{\delta\epsilon\kappa}-\frac{3}{2}\mathcal{D}_\alpha Q^{\alpha\gamma\delta}Q_{\gamma\epsilon\kappa}Q_\delta{}^{\epsilon\kappa}+\mathcal{D}_\mu \Bar{Q}_\nu Q^\mu Q^\nu +\frac{1}{2}\mathcal{D}_\mu Q^{\mu\nu\rho}Q_\nu Q_\rho \\
	&+2\mathcal{D}_\alpha \Bar{Q}^\delta Q^{\alpha\epsilon\kappa}Q_{\epsilon\delta\kappa}-2\mathcal{D}_\alpha \Bar{Q}_{\delta}Q^{\alpha\delta\epsilon}Q_\epsilon+\mathcal{D}_\alpha Q^{\alpha\gamma\delta}Q_{\gamma\delta\epsilon}Q^\epsilon\\
	&-\frac{1}{2}\mathcal{D}_\mu Q_\nu Q^\mu Q^\nu+\frac{1}{2}\mathcal{D}_\alpha Q^{\beta}Q^{\alpha\epsilon\kappa}Q_{\beta\epsilon\kappa}-\mathcal{D}_\alpha Q^{\beta}Q^{\alpha\epsilon\kappa}Q_{\epsilon\beta\kappa}\\
	&-4\mathcal{D}_\alpha Q_{\beta\gamma\delta}Q^{\delta\gamma\kappa}Q_\kappa{}^{\alpha\beta}+6\mathcal{D}_\alpha Q_{\beta\gamma\delta}Q^{\beta\gamma\kappa}Q_\kappa{}^{\alpha\delta}+6\mathcal{D}_\alpha Q_{\beta\gamma\delta}Q^{\kappa\alpha\beta}Q_\kappa{}^{\gamma\delta}\\
	&-\mathcal{D}_{\beta} Q_{\alpha\gamma\delta}Q^{\beta\gamma\mu}Q^{\delta\alpha}{}_\mu+\frac{1}{2}\mathcal{D}_\alpha Q_{\beta\gamma\delta}Q^{\mu\alpha\gamma}Q_\mu{}^{\beta\delta},
	\end{split}
	\end{align}
	with 21 terms. The $\mathcal{G}_{\mathcal{DD}Q}$ invariant explicitly reads
	\begin{align}
	\begin{split}
	\mathcal{G}_{\mathcal{DD}Q}&=-\frac{1}{2}\mathcal{D}_\alpha  Q_{\beta\kappa\mu}\mathcal{D}^\alpha Q^{\kappa\beta\mu}+\frac{3}{4}\mathcal{D}_\alpha Q_{\kappa\beta\mu}\mathcal{D}^\alpha Q^{\kappa\beta\mu}-\mathcal{D}_\mu Q_\nu\mathcal{D}^\mu Q^\nu-\mathcal{D}_\nu Q^\nu{}_{\sigma\mu}\mathcal{D}_\rho Q^{\rho\sigma\mu}\\
	&+4\mathcal{D}_\nu Q^\nu{}_{\kappa\lambda}\mathcal{D}_\mu Q^{\kappa\lambda\mu}-2\mathcal{D}_\mu Q^\mu \mathcal{D}_\nu \Bar{Q}^\nu-2\mathcal{D}_\nu Q^{\nu\rho}{}_{\kappa}\mathcal{D}_\rho Q^\kappa+\mathcal{D}_\mu Q^\mu \mathcal{D}_\nu Q^\nu\\
	&-2\mathcal{D}_\nu Q^{\beta\nu\kappa}\mathcal{D}_\mu Q_{\kappa\beta}{}^\mu+\mathcal{D}_\mu \Bar{Q}^\mu \mathcal{D}_\nu \Bar{Q}^\nu+2\mathcal{D}_\mu Q^{\nu\mu\rho}\mathcal{D}_\nu Q_\rho-2\mathcal{D}_\alpha Q^{\beta\alpha\lambda}\mathcal{D}_\mu Q_{\beta\lambda}{}^\mu\\
	&+2\mathcal{D}_\mu Q^{\rho\mu\nu}\mathcal{D}_\nu Q_\rho.
	\end{split}
	\end{align}
	with 13 terms.
	
	Finally, we present the Gauss-Bonnet invariant written in the general telaparallel case, starting with $\mathcal{G}_{TTTQ}$:
	\begin{align}
	\begin{split}
	\mathcal{G}_{TTTQ}&=\Bar{Q}_\mu T^\mu T_\nu T^\nu-\Bar{Q}_\mu T^\mu T_{\nu\sigma\rho}T^{\sigma\nu\rho}-\frac{1}{4}\Bar{Q}_\mu T^\mu T_{\rho\nu\sigma}T^{\rho\nu\sigma}\\
	&+\frac{1}{4}Q_{\alpha\delta\mu}T^{\alpha\delta\nu}T^{\mu\sigma\rho}T_{\nu\sigma\rho}+Q_{\alpha\delta\mu}T^{\delta\mu\rho}T^{\nu\alpha\sigma}T_{\nu\sigma\rho}+\frac{1}{4}Q_{\alpha\delta\mu} T^{\delta\alpha\mu}T_{\nu\sigma\rho}T^{\nu\sigma\rho}\\
	&-2Q_{\alpha\delta\mu}T^{\nu\alpha}{}_\rho T_{\nu}{}^{\delta\rho}T^\mu+2Q_{\alpha\delta\mu}T^{[\delta\alpha]\nu}T^\mu{}_{\nu\rho}T^\rho+2Q_{\alpha\delta\mu}T^{\delta\mu\rho}T^\alpha T_\rho\\
	&-Q_{\alpha\delta\mu}T^{\delta\alpha\mu}T_\rho T^\rho+2Q_{\alpha\delta\mu}T^{\delta\mu\nu}T_\nu{}^{\alpha\rho}T_\rho-Q_{\alpha\delta\mu}T^{\alpha\delta}{}_\nu T^{\nu\mu\rho}T_{\rho}\\
	&+2Q_{\alpha\delta\mu}T_\nu{}^{\alpha\delta}T^{\nu\mu\rho}T_\rho-Q_{\alpha\delta\mu}T^{\delta\nu\sigma}T^\mu{}_{\nu\rho}T_\sigma{}^{\alpha\rho}-\frac{3}{2}Q_{\alpha\delta\mu}T^{\delta\mu\nu}T_{\nu\sigma\rho}T^{\sigma\alpha\rho}\\
	&+\frac{1}{4}Q_{\alpha\delta\mu}T_\nu{}^{\alpha\rho}T^{\nu\delta}{}_\sigma T^{\sigma\mu}{}_\rho-\frac{7}{4}Q_{\alpha\delta\mu}T^{\nu\alpha}{}_\sigma T_\nu{}^{\delta\rho}T^{\sigma\mu}{}_\rho-4Q_{\alpha\delta\mu}T^{(\alpha\delta)\rho}T^\mu T_\rho\\
	&-Q_{\alpha\delta\mu}T^{\nu\alpha\delta}T_{\nu\sigma\rho}T^{\sigma\mu\rho}+\frac{3}{4}Q_{\alpha\delta}{}^\mu T^{\nu\alpha\rho}T^{\sigma\delta}{}_\nu T_{\sigma\mu\rho}-\frac{1}{2}Q_{\alpha\delta\mu}T^{\alpha}T_\sigma{}^{\delta\rho}T^{\sigma\mu}{}_{\rho}\\
	&-\frac{1}{2}Q_{\alpha\delta\mu}T^{\delta\alpha\nu}T^{\mu\sigma\rho}T_{\sigma\nu\rho}-\frac{1}{2}Q_{\alpha\delta\mu}T^{\alpha\delta\rho}T^{\nu\mu\sigma}T_{\sigma\nu\rho}+\frac{1}{2}Q_{\alpha\delta\mu}T^{\delta\alpha\mu}T^{\nu\sigma\rho}T_{\sigma\nu\rho}\\
	&+\frac{1}{2}Q_{\alpha\delta\mu}T^{\nu\alpha\delta}T^{\sigma\mu\rho}T_{\sigma\nu\rho}-\frac{1}{2}Q_{\alpha\delta\mu}T^{\nu\delta\sigma}T_{\sigma\nu\rho}T^{\rho\alpha\mu}-Q_{\alpha\delta\mu}T^{\delta\mu\nu}T_\rho T^{\rho\alpha}{}_\nu\\
	&+\frac{1}{2}Q_{\alpha\delta\mu}T^{\alpha\delta\nu}T^{\sigma\mu\rho}T_{\sigma\nu\rho}+\frac{1}{4}Q_{\alpha\delta\mu}T^{\nu\delta}{}_\sigma T_\nu{}^{\mu\rho}T_\rho{}^{\alpha\sigma}-\frac{1}{2}Q_{\alpha\delta\mu}T^{\delta\nu\rho}T^{\mu}{}_{\nu\rho}T^\alpha\\
	&+2Q_{[\mu\alpha]\delta}T^{\delta\nu\rho}T_{\nu}{}^{\mu\rho}T^\alpha +2Q_{\alpha\delta\mu}T^{\delta\nu\rho}T_\nu{}^{\alpha\mu}T_\rho-Q_{\alpha\delta\mu}T^{\delta\nu\rho}T_{\nu\sigma\rho}T^{\sigma\alpha\mu}\\
	&+\frac{7}{4}Q_{\alpha\delta}{}^\mu T^{\delta\nu\rho}T^{\sigma\alpha}{}_\rho T_{\sigma\mu\nu}+Q_{\alpha\delta\mu}T^{\delta\nu\rho}T_{\nu\sigma}{}^\alpha T^{\sigma\mu}{}_\rho \\
	&-\frac{3}{4}Q_{\alpha\delta\mu}T^{\delta\nu\rho}T^{\sigma\alpha\mu}T_{\sigma\nu\rho}+2Q_{\alpha\delta\mu}T^{[\delta\alpha]}{}_\nu T_\rho T^{\rho\mu\nu}+Q_{\alpha\delta}{}^\mu T^{\alpha\sigma\nu}T^{\delta\rho}{}_\nu T_{\rho\mu\sigma}\\
	&+Q_{\alpha\delta\mu}T^{\alpha}{}_{\nu\sigma} T^{\nu\delta\rho}T_\rho{}^{\mu\sigma}+Q_{\alpha\delta\mu}T^{\alpha\delta\nu}T^{\mu\sigma\rho}T_{\rho\nu\sigma}-Q_{\alpha\delta\mu}T^{\alpha\nu\sigma}T^{\delta\mu\rho}T_{\rho\nu\sigma}\\
	&-\frac{1}{4}Q_{\alpha\delta\mu}T^{\delta\alpha\rho}T^{\mu\nu\sigma}T_{\rho\nu\sigma}+Q_{\alpha\delta\mu}T^{\delta\alpha\nu}T^{\mu\sigma\rho}T_{\rho\nu\sigma}+\frac{1}{2}Q_{\alpha\delta\mu}T^{\delta\alpha\nu}T^{\sigma\mu\rho}T_{\rho\nu\sigma}\\
	&+2Q_{\alpha\delta\mu}T^\alpha{}_{\nu\rho}T^{(\delta\nu)\rho}T^\mu+Q_{\alpha\delta\mu}T^{\alpha\nu\rho}T^{\delta\mu}{}_\nu T_\rho-\frac{1}{4}Q_{\alpha\delta\mu}T^{\alpha\nu\rho}T_{\sigma\nu}{}^\delta T^{\sigma\mu}{}_\rho \\
	&-\frac{3}{4}Q_\mu T^\mu T_{\nu\sigma\rho}T^{(\nu\sigma)\rho}+\frac{1}{4}Q_{\alpha\delta\mu}T^{\alpha\nu\rho}T^{\delta}{}_{\nu\sigma}T^{\mu\sigma}{}_\rho-\frac{1}{4}Q_{\alpha\delta\mu}T^{\alpha}{}_{\nu\rho} T^{\nu\delta}{}_\sigma T^{\rho\mu\sigma}\\
	&-\frac{1}{4}Q^\alpha{}_{\delta\mu}T^{\nu\delta\sigma}T^{\rho\mu}{}_\nu T_{\rho\alpha\sigma}-\frac{1}{2}Q_{\alpha\delta\mu}T^\alpha T^{\rho\mu}{}_\sigma T_{\rho}{}^{\delta\sigma}+\frac{1}{2}Q_{\alpha\delta\mu}T^{\delta\alpha\nu}T_{\rho\nu\sigma}T^{\rho\mu\sigma}\\
	&+\frac{1}{2}Q_\alpha T^{\alpha\mu \nu}T^\rho T_{\rho\mu\nu} +Q_\alpha T^{\alpha\mu}{}_\nu T_{(\mu\sigma)\rho}T^{\rho\nu\sigma}-2Q_\alpha T^{\mu\alpha\nu}T_{(\mu\nu)\rho}T^\rho\\
	&+\frac{1}{2}Q_\alpha T^{\mu\alpha\nu}T^\sigma{}_{\mu\rho}T_{\nu\sigma}{}^\rho +Q_\alpha T^{\mu\alpha}{}_\nu T_{\mu\sigma\rho}T^{(\nu\sigma)\rho}+Q_\mu T^\mu T_\nu T^\nu\\
	&-\frac{3}{4}Q_\mu T^\mu T_{\nu\sigma\rho}T^{(\nu\sigma)\rho}
	\end{split}
	\end{align}
	with 70 terms. Regarding the $\mathcal{G}_\mathrm{TTQQ}$-term the result is lengthy, and for this reason we split it into two terms $\mathcal{G}_{TTQQ}=\overset{1}{\mathcal{G}}_{TTQQ}+\overset{2}{\mathcal{G}}_{TTQQ}$
	\begin{align}
	\begin{split}
	\overset{1}{\mathcal{G}}_{TTQQ}&=\frac{1}{2}Q_{\alpha\gamma\epsilon}Q^{\alpha\gamma\kappa}T^{\epsilon\mu\rho}T_{\kappa\mu\rho}-\frac{1}{4}Q_{\alpha\gamma\epsilon}Q^{\gamma\alpha\kappa}T^{\epsilon\mu\rho}T_{\kappa\mu\rho}+\frac{1}{2}Q_{\alpha\gamma\epsilon}Q^{\kappa\gamma\epsilon}T^{\alpha}{}_{\kappa\rho}T^\rho\\
	&+2Q_{\alpha\gamma\epsilon}Q^{[\gamma\alpha]\kappa}T^\epsilon{}_{\kappa\rho}T^\rho-\frac{1}{4}Q_{\alpha\gamma\epsilon}Q^{\kappa\gamma\epsilon}T^{\alpha\mu\rho}T_{\mu\kappa\rho}+Q_{\alpha\gamma\epsilon}Q^{[\alpha\gamma]\kappa}T^{\epsilon\mu\rho}T_{\mu\kappa\rho}\\
	&-\frac{3}{2}Q_{\alpha\gamma\epsilon}Q^{\kappa\gamma\mu}T^{\epsilon}{}_{\kappa\rho}T^{\rho\alpha}{}_\mu+\frac{1}{2}Q_{\alpha\gamma\epsilon}Q^{\kappa\gamma\mu}T^{\alpha}{}_{\mu\rho}T^{\rho\epsilon}{}_\kappa+2Q_{\alpha\gamma\epsilon}Q^{\kappa\gamma\mu}T^{\alpha\epsilon\rho}T_{\rho\kappa\mu}\\
	&-2Q_{\alpha\gamma\epsilon}Q^{\gamma\kappa\mu}T^{(\alpha\epsilon)\rho}T_{\kappa\mu\rho}+\frac{3}{4}Q_{\alpha\gamma\epsilon}Q^{\gamma\kappa\mu}T^\epsilon{}_{\kappa\rho}T_\mu{}^{\alpha\rho}+\frac{3}{4}Q_{\alpha\gamma\epsilon}Q^{\gamma\kappa\mu}T_\kappa{}^{\alpha\rho} T_{\mu}{}^{\epsilon}{}_\rho\\
	&-2Q_{\alpha\gamma\epsilon}Q^{\gamma\kappa\mu}T_{\kappa\mu\rho}T^{\rho\alpha\epsilon}+\frac{3}{4}Q_{\alpha\gamma}{}^\epsilon Q^{\gamma\kappa\mu}T_{\kappa\epsilon\rho}T^{\rho\alpha}{}_\mu+\frac{3}{2}Q_{\alpha\gamma\epsilon}Q^{\gamma\kappa}{}_\mu T^\alpha{}_{\kappa\rho}T^{(\epsilon\mu)\rho}\\
	&+\frac{3}{4}Q_{\alpha\gamma}{}^\epsilon Q^{\gamma\kappa\mu}T_{\epsilon\kappa\rho}T^{\rho\alpha}{}_\mu-\frac{3}{2}Q^\alpha{}_{\gamma\epsilon}Q^{\gamma\kappa\mu}T_{(\alpha\kappa)\rho}T^{\rho\epsilon}{}_\mu-\frac{3}{4}Q^\alpha{}_{\gamma\epsilon}Q^{\gamma\kappa\mu}T^{\rho\epsilon}{}_\mu T_{\rho\alpha\kappa}\\
	&-Q_{\alpha\gamma}{}^\epsilon Q^{\alpha\kappa\mu}T^\gamma{}_{\kappa\rho}T^{\epsilon\mu\rho}-\frac{1}{2}Q_{\alpha\gamma}{}^\epsilon Q^{\alpha\kappa\mu}T^\gamma{}_{\kappa\rho}T^{\mu\epsilon\rho}+\frac{1}{2}Q_{\alpha\gamma\epsilon}Q^{\alpha\kappa\mu}T^{\gamma\epsilon\rho}T_{\kappa\mu\rho}\\
	&+\frac{1}{4}Q_{\alpha\gamma\epsilon}Q^{\alpha\kappa\mu}T^{\rho\epsilon}{}_\mu T_\rho{}^\gamma{}_\kappa-Q_\alpha Q^{\epsilon\alpha\kappa}T_{(\epsilon\kappa)\rho}T^\rho +\frac{1}{4}Q_\alpha Q^{\epsilon\alpha\kappa}T_{\kappa\mu}{}^\rho T^\mu{}_{\epsilon\rho}\\
	&+\frac{1}{2}Q_\alpha Q^{\epsilon\alpha}{}_\kappa T_{\epsilon\mu\rho}T^{(\kappa\mu)\rho}+\frac{1}{2}Q_\alpha Q^{\alpha\epsilon\kappa}T_{\epsilon\kappa\rho}T^\rho-\frac{1}{8}Q_\alpha Q^{\alpha\epsilon\kappa}T_{\epsilon\mu\rho}T_\kappa{}^{\mu\rho}\\
	&-\frac{1}{4}Q_\alpha Q^{\alpha\epsilon\kappa}T_{\epsilon\mu}{}^\rho T^{\mu}{}_{\kappa\rho}-2\Bar{Q}_\alpha Q_{\epsilon\kappa\mu}T^{\kappa\epsilon\mu}T^\alpha+2\Bar{Q}_\mu Q_{\nu}T^\mu T^\nu\\
	&-\Bar{Q}_\mu \Bar{Q}_\nu T^\mu T^\nu+Q_\alpha Q_{\epsilon\kappa\mu}T^{(\alpha\epsilon)\rho}T^{\kappa\mu}{}_\rho-\frac{1}{2}Q_\alpha Q_{\epsilon\kappa\mu}T^{\epsilon\kappa\rho}T^{\mu\alpha}{}_\rho+Q_\alpha Q_{\epsilon\kappa\mu}T^{\alpha\kappa\rho}T^{[\epsilon\mu]}{}_\rho\\
	&+\frac{1}{2}Q_\alpha Q_{\epsilon\kappa\mu}T^{\kappa\alpha}{}_\rho T^{\mu\epsilon\rho}+2Q_{(\alpha} Q_{\epsilon)\kappa\mu}T^{\kappa\epsilon\mu}T^\alpha+2Q_\alpha Q_{\epsilon\kappa\mu}T^{(\epsilon\kappa)\alpha}T^\mu\\
	&+\frac{1}{2}Q_\alpha Q_{\epsilon\kappa\mu}T^{\kappa\mu}{}_\rho T^{\rho\alpha\epsilon}+Q_\alpha Q_{\epsilon\kappa\mu}T^{[\kappa\epsilon]}{}_\rho T^{\rho\alpha\mu}+Q_\alpha Q_{\epsilon\kappa\mu}T^{\kappa\alpha\rho}T_\rho{}^{\epsilon\mu}\\
	&+Q_\alpha Q_{\epsilon\kappa\mu}T^{\rho\alpha\kappa}T_\rho{}^{\epsilon\mu}+\frac{1}{8}Q_\alpha Q_\epsilon T^{\alpha\mu\rho} T^{\epsilon}{}_{\mu\rho}-\frac{1}{2}Q_\mu Q_\nu T^\mu T^\nu-\frac{1}{2}Q_\alpha Q_\epsilon T^{\mu\alpha\rho} T_{(\mu\rho)}{}^\epsilon\\
	&-\frac{1}{2}Q_{\alpha\gamma\epsilon}Q^{\kappa\gamma\mu}T^{\alpha}{}_{\kappa\rho}T^{\epsilon}{}_\mu{}^\rho-\frac{1}{2}Q_{\alpha\gamma\epsilon}Q^{\kappa\gamma\mu}T^{\alpha}{}_{\mu\rho}T_\kappa{}^{\epsilon\rho}-\frac{1}{2}Q_{\alpha\gamma\epsilon}Q^{\kappa\gamma\mu}T^{\alpha\epsilon\rho}T_{\kappa\mu\rho}\\
	&-\frac{1}{4}Q_{\alpha\gamma\epsilon}Q^{\gamma\epsilon\kappa}T^{\alpha\mu\rho}T_{\kappa\mu\rho}+Q_{[\alpha\epsilon]\gamma}Q^{\kappa\gamma\mu}T^{\alpha}{}_{\kappa\rho}T_\mu{}^{\epsilon\rho}+Q_{\alpha\gamma\epsilon}Q^{\kappa\gamma\mu}T^{\alpha\epsilon\rho}T_{\mu\kappa\rho}\\
	&-\frac{1}{2}Q_{\alpha\gamma\epsilon}Q^{\kappa\gamma\mu}T^{\epsilon\alpha\rho}T_{\mu\kappa\rho}-\frac{1}{2}Q_{\alpha\gamma\epsilon}Q^{\gamma\epsilon\kappa}T^{\mu\alpha\rho}T_{\mu\kappa\rho}+\frac{1}{4}Q_{\alpha\gamma\epsilon}Q^{\kappa\gamma\epsilon}T^{\mu\alpha\rho}T_{\mu\kappa\rho}\\
	&+\frac{1}{2}Q_{\alpha\gamma\epsilon}Q^{\alpha\gamma\kappa}T^{\mu\epsilon\rho}T_{\mu\kappa\rho}+4Q_{[\alpha\epsilon]\gamma}Q^{\kappa\gamma\mu}T^{\alpha}{}_{\kappa\mu}T^\epsilon+2Q_{\alpha\gamma\epsilon}Q^{\kappa\gamma\mu}T^{[\epsilon\alpha]}{}_\mu T_\kappa\\
	&+2Q_{\alpha\gamma\epsilon}Q^{\gamma\epsilon\kappa}T^\alpha T_\kappa-\frac{1}{2}Q_{\alpha\gamma\epsilon}Q^{\kappa\gamma\epsilon}T^\alpha T_\kappa
	\end{split}
	\end{align}
	\begin{align}
	\begin{split}
	\overset{2}{\mathcal{G}}_{TTQQ}  &=-2Q_{\alpha\gamma\epsilon}Q^{(\alpha\gamma)\kappa}T^\epsilon T_\kappa+2Q_{\alpha\gamma\epsilon}Q^{\kappa\gamma\mu}T^{[\epsilon\alpha]}{}_\kappa T_\mu+Q_{\alpha\gamma\epsilon}Q^{\alpha\kappa\mu}T^{\gamma\epsilon}{}_\kappa T_\mu\\
	&-\frac{3}{2}Q_{\alpha\gamma\epsilon}Q^{\kappa\mu\rho}T^\alpha{}_{\kappa\mu}T^{\gamma\epsilon}{}_\rho+\frac{3}{2}Q_{\alpha\gamma\epsilon}Q^{\kappa\mu\rho}T^{\alpha\gamma}{}_\mu T^\epsilon{}_{\kappa\rho}-\frac{1}{2}Q_{\alpha\gamma\epsilon}Q^{\kappa\mu\rho}T^{\gamma\alpha}{}_\mu T^\epsilon{}_{\kappa\rho}\\
	&-\frac{1}{2}Q_{\alpha\gamma\epsilon}Q^{\kappa\mu\rho}T^{\gamma\epsilon}{}_\kappa T_\mu{}^\alpha{}_\rho -\frac{1}{2}Q_{\alpha\gamma\epsilon}Q^{\kappa\mu}{}_\rho T^{\alpha\gamma}{}_\kappa T_\mu{}^{\epsilon\rho}+\frac{3}{2}Q_{\alpha\gamma\epsilon}Q^{\kappa\mu}{}_\rho T^{\gamma\alpha}{}_\kappa T_\mu{}^{\epsilon\rho}\\
	&+\frac{1}{2}Q_{\alpha\gamma\epsilon}Q^{\kappa\mu\rho}T^{\gamma\alpha\epsilon}T_{\mu\kappa\rho}-Q_{\alpha\gamma\epsilon}Q^{\kappa\mu\rho}T^\gamma{}_{\kappa\mu}T_\rho{}^{\alpha\epsilon}+\frac{1}{2}Q_{\alpha\gamma\epsilon}Q^{\kappa\gamma\mu}T^\epsilon{}_{\mu\rho}T^{\rho\alpha}{}_\kappa\\
	&+\frac{1}{2}Q_{\alpha\gamma\epsilon}Q^{\kappa\gamma\mu}T^{\rho\alpha}{}_\mu T_{\rho}{}^\epsilon{}_\kappa+\frac{1}{2}Q_{\alpha\gamma\epsilon}Q^{\kappa\gamma\mu}T^\alpha{}_{\kappa\rho}T^{\rho\epsilon}{}_\mu +\frac{1}{2}Q_{\alpha\gamma\epsilon}Q^{\alpha\kappa\mu}T^\gamma{}_{\kappa\rho}T^{\rho\epsilon}{}_\mu\\
	&-\frac{1}{4}Q_{\alpha\gamma\epsilon}Q^{\kappa\gamma\mu}T^{\rho\alpha}{}_\kappa T_\rho{}^\epsilon{}_\mu+2Q_{[\alpha\epsilon]\gamma}Q^{\gamma\kappa\mu}T_\kappa{}^\epsilon{}_\mu T^\alpha+2Q_{\alpha\gamma\epsilon}Q^{\gamma\kappa\mu}T^{[\epsilon\alpha]}{}_\kappa T_\mu\\
	&+2Q_{\alpha\gamma\epsilon}Q^{\gamma\kappa\mu}T_\kappa{}^{\alpha\epsilon}T_\mu-\frac{1}{2}Q_{\alpha\gamma\epsilon}Q^{\gamma\epsilon\kappa}T^{\mu\alpha\rho}T_{\rho\kappa\mu}+\frac{1}{4}Q_{\alpha\gamma\epsilon}Q^{\kappa\gamma\epsilon}T^{\mu\alpha\rho}T_{\rho\kappa\mu}\\
	&+\frac{1}{2}Q_{\alpha\gamma\epsilon}Q^{\alpha\gamma\kappa}T^{\mu\epsilon\rho}T_{\rho\kappa\mu}-Q_{\alpha\gamma\epsilon}Q^{\kappa\gamma\mu}T^{\rho\alpha\epsilon}T_{\rho\kappa\mu}.
	\end{split}
	\end{align}
	with 73+27=100 terms. Then, $\mathcal{G}_{TQQQ}$ explicitly reads
	\begin{align}
	\begin{split}
	\mathcal{G}_{TQQQ}&=\frac{1}{4}\Bar{Q}_\mu Q^\nu Q_\nu T^\mu-\frac{1}{4}\Bar{Q}_\mu Q^{\sigma \nu\rho}Q_{\sigma\nu\rho}T^\mu+\frac{1}{2}\Bar{Q}_\mu Q^{\sigma\nu\rho}Q_{\nu\sigma\rho}T^\mu+\frac{1}{2}\Bar{Q}_\mu Q^\mu Q^{\nu[\rho\sigma]}T_{\rho\nu\sigma}\\
	&+\frac{1}{2}\Bar{Q}_\mu Q^\mu Q_\nu T^\nu-\frac{1}{2}\Bar{Q}_\mu Q^\mu \Bar{Q}_\nu T^\nu-\frac{1}{2}Q_{\alpha\gamma\epsilon}Q^{\gamma\epsilon\mu}Q_{\rho\mu\tau}T^{\alpha\rho\tau}+Q_{\alpha\gamma\epsilon}Q^{\alpha\mu\rho}Q^{\tau\gamma}{}_\mu T^\epsilon{}_{\rho\tau}\\
	&+\frac{1}{2}Q_{\alpha\gamma\epsilon}Q^{\alpha\mu\rho}Q^{\gamma\epsilon\tau}T_{\mu\rho\tau}-\frac{1}{2}Q_{\alpha\gamma\epsilon}Q^{\gamma\epsilon\mu}Q^{\rho\alpha\tau}T_{\mu\rho\tau}-\frac{3}{2}Q_{\alpha\gamma\epsilon}Q^{\alpha\gamma\mu}Q^{\rho\epsilon\tau}T_{\mu\rho\tau}\\
	&+\frac{5}{2}Q_{\alpha\gamma\epsilon}Q^{\gamma\alpha\mu}Q^{\rho\epsilon\tau}T_{\mu\rho\tau}-\frac{1}{2}Q_{\alpha\gamma\epsilon}Q^{\gamma\epsilon\mu}Q_{\mu\rho\tau}T^{\rho\alpha\tau}-Q_{\alpha\gamma\epsilon}Q^{\alpha\mu\rho}Q^\gamma{}_{\mu\tau}T_\rho{}^{\epsilon\tau}\\
	&+2Q_{\alpha\gamma\epsilon}Q^{\alpha\gamma\mu}Q^{\epsilon\rho\tau}T_{\rho\mu\tau}-Q_{\alpha\gamma\epsilon}Q^{\gamma\alpha\mu}Q^{\epsilon\rho\tau}T_{\rho\mu\tau}+\frac{1}{4}Q_{\alpha\gamma\epsilon}Q^{\alpha\gamma\epsilon}Q^{\mu\rho\tau}T_{\rho\mu\tau}\\
	&-\frac{1}{2}Q_{\alpha\gamma\epsilon}Q^{\gamma\alpha\epsilon}Q^{\mu\rho\tau}T_{\rho\mu\tau}+3Q_{\alpha\gamma\epsilon}Q^{[\gamma\alpha]\mu}Q^{\rho\epsilon\tau}T_{\rho\mu\tau}-Q_{\alpha\gamma\epsilon}Q^{\alpha\mu\rho}Q^\gamma{}_{\mu\rho}T^\epsilon\\
	&+4Q_{\alpha\gamma\epsilon}Q^{[\alpha\gamma]\mu}Q^\epsilon{}_{\mu\rho}T^\rho+2Q_{\alpha\gamma\epsilon}Q^{[\gamma\alpha]\mu}Q^{\rho\epsilon}{}_\mu T_\rho+\frac{1}{2}Q_{\alpha\gamma\epsilon}Q^{\gamma\mu\rho}Q^\epsilon{}_{\mu\rho}T^\alpha\\
	&-\frac{1}{4}Q_{\alpha\gamma\epsilon}Q^{\gamma\mu\rho}Q_{\tau\mu\rho}T^{\alpha\epsilon\tau}+\frac{3}{2}Q_{\alpha\gamma\epsilon}Q^{\gamma\mu\rho}Q_{\tau\mu\rho}T^{\epsilon\alpha\tau}+\frac{5}{2}Q_{\alpha\gamma\epsilon}Q^{\gamma\mu\rho}Q_{\tau\mu\rho}T^{\tau\alpha\epsilon}\\
	&+Q_{\alpha\gamma\epsilon}Q^{\gamma\mu}{}_\rho Q_\mu{}^{\epsilon\tau}T^{(\alpha\rho)}{}_\tau+\frac{1}{2}Q_{\alpha\gamma\epsilon}Q^{\gamma\mu\rho}Q_\mu{}^{\epsilon\tau}T_\tau{}^\alpha{}_\rho-Q_{\alpha\gamma\epsilon}Q^{\gamma\mu}{}_\rho Q^{\tau\epsilon}{}_\mu T^{(\alpha\rho)}{}_\tau\\
	&-\frac{1}{2}Q_{\alpha\gamma\epsilon}Q^{\gamma\mu}{}_\rho Q^{\tau\epsilon}{}_\mu T_\tau{}^{\alpha\rho}+Q_{\alpha\gamma\epsilon}Q^{[\alpha\gamma]\mu}Q^{\rho\epsilon\tau}T_{\tau\mu\rho}-\frac{1}{2}Q_\alpha Q^{\epsilon\alpha\mu}\Bar{Q}^\rho T_{\epsilon\mu\rho}\\
	&-Q_\alpha Q^{\epsilon\alpha\mu}Q_{\rho\mu\tau}T_\epsilon{}^{\rho\tau}-\frac{1}{2}Q_\alpha Q^{\epsilon\alpha\mu}Q^\rho T_{\mu\epsilon\rho}-Q_\alpha Q^{\epsilon\alpha\mu}Q_{\rho\epsilon\tau}T_\mu{}^{\rho\tau}\\
	&+Q_\alpha Q^{\epsilon\alpha\mu}Q_{[\mu\rho]}{}^\tau T^\rho{}_{\epsilon\tau}+Q_\alpha Q^{\epsilon\alpha\mu}Q_{[\epsilon\rho]}{}^\tau T^\rho{}_{\mu\tau}-\frac{1}{4}Q_\mu Q^\mu Q_{\nu\rho\sigma}T^{\rho\nu\sigma}\\
	&-\frac{1}{2}Q_\alpha Q_{\epsilon\mu\rho}Q^{\tau\mu\rho}T^{\epsilon\alpha}{}_\tau+2Q_\alpha Q_{\epsilon\mu\rho} Q^{[\epsilon\mu]\tau}T^{\rho\alpha}{}_\tau+\frac{1}{4}Q_\alpha Q_{\epsilon\mu\rho} Q^{\epsilon\mu\rho}T^\alpha\\
	&-\frac{1}{2}Q_\alpha Q_{\epsilon\mu\rho} Q^{\mu\epsilon\rho}T^\alpha-\frac{1}{4}Q_\mu Q^\mu Q_\nu T^\nu +2Q_\alpha Q^{\epsilon\alpha\mu}T_{\epsilon\mu\rho}T^\rho -Q^\alpha Q_{\epsilon\alpha\mu}Q^{\rho\epsilon\mu}T_\rho\\
	&+\frac{1}{2}Q_\alpha Q^{\epsilon\alpha\mu}Q^\rho{}_{\mu\tau}T^{\tau}{}_{\epsilon\rho}+\frac{1}{2}Q_\alpha Q^{\alpha\epsilon\mu}Q^\rho T_{\epsilon\mu\rho}-\frac{1}{2}Q_\alpha Q^{\alpha\epsilon}{}_\mu Q_{\epsilon\rho\tau}T^{\rho\mu\tau}\\
	&-Q_\alpha Q^{\alpha\epsilon\mu}Q_{\epsilon\mu\rho}T^\rho+\frac{1}{2}Q_\alpha Q^{\alpha\epsilon\mu} Q_{\rho\epsilon\mu}T^\rho+Q_\alpha Q^{\alpha\epsilon\mu}Q_{\rho\epsilon\tau}T_\mu{}^{\rho\tau}+Q_\alpha Q^{\alpha\epsilon\mu}Q_{\rho\epsilon\tau}T^{[\tau\rho]}{}_\mu\\
	&+\frac{1}{2}Q_\alpha Q^{\epsilon\alpha}{}_\mu Q_{\rho\epsilon\tau}T^{\tau\mu\rho},
	\end{split}
	\end{align}
	with 65 terms. Next, we move on to terms with higher order derivatives, starting with $\mathcal{G}_{TQ\mathcal{D}T}$
	\begin{align}
	\begin{split}
	\mathcal{G}_{TQ\mathcal{D}T}&=4\mathcal{D}_\alpha T^{\gamma\alpha\epsilon}Q_\mu T_{(\gamma\epsilon)}{}^\mu +8\mathcal{D}_\alpha T^{\gamma\alpha\epsilon}\Bar{Q}_\mu T_{(\gamma\epsilon)}{}^\mu-4\mathcal{D}_\alpha T^{\gamma\alpha\epsilon}Q_{(\epsilon\mu)}{}^\rho T^\mu{}_{\gamma\rho}\\
	&-4\mathcal{D}_\alpha T^{\gamma\alpha\epsilon}Q_{(\gamma\mu)}{}^\rho T^\mu{}_{\epsilon\rho}+8\mathcal{D}_\alpha T^{\gamma\alpha\epsilon}Q_{(\gamma\epsilon)\mu}T^\mu+4\mathcal{D}_\alpha T^{\gamma\alpha\epsilon}Q_{\mu\gamma\epsilon}T^\mu\\
	&+4\mathcal{D}_\alpha T^{(\gamma\epsilon)\alpha}Q_{\mu\epsilon\rho}T^{\rho}{}_\gamma{}^\mu-4\mathcal{D}_\alpha T_\epsilon Q_\mu T^{(\alpha\epsilon)\mu}+2\mathcal{D}_\alpha T_\epsilon Q_\mu T^{\mu\alpha\epsilon}-4\mathcal{D}_\alpha T_\epsilon \Bar{Q}_\mu T^{(\alpha\epsilon)\mu}\\
	&+2\mathcal{D}_\alpha T_\epsilon \Bar{Q}_\mu T^{\mu\alpha\epsilon}+4\mathcal{D}_\alpha T^\epsilon Q^{\alpha\mu\rho}T_{\mu\epsilon\rho}-8\mathcal{D}_\alpha T_\epsilon Q^{(\alpha\epsilon)\mu}T_\mu -4\mathcal{D}_\alpha T_\epsilon Q^{\mu\alpha\epsilon}T_\mu \\
	&-8 \mathcal{D}_\alpha T^\epsilon Q^{\mu\alpha\rho}T_{(\mu\rho)\epsilon}-\mathcal{D}_\alpha T^{\alpha\delta\epsilon} Q^\mu T_{\mu\delta\epsilon}-4\mathcal{D}_\alpha T^{\alpha\delta\epsilon} Q_{(\delta\mu)}{}^\rho T^{\mu}{}_{\epsilon\rho}-2\mathcal{D}_\alpha T^{\alpha\delta\epsilon} \Bar{Q}^\mu T_{\mu \delta\epsilon}\\
	&-2\mathcal{D}_\alpha T^{\alpha\delta}{}_\epsilon Q_{\mu\delta\rho}T^{\rho\epsilon\mu},
	\end{split}
	\end{align}
	with 30 terms. Explicity $\mathcal{G}_{QQ\mathcal{D}T}$ reads
	\begin{align}
	\begin{split}
	\mathcal{G}_{QQ\mathcal{D}T}&=\mathcal{D}_\alpha T_{\gamma\delta\epsilon}Q^{(\alpha\gamma)\rho}Q^{\delta\epsilon}{}_\rho+\frac{1}{2}\mathcal{D}_\alpha T^{\gamma\delta\epsilon} Q_{\gamma\delta\rho}Q_\epsilon{}^{\alpha\rho}-\frac{1}{2}\mathcal{D}_\alpha T^{\gamma\delta\epsilon} Q_{\delta}{}^{\alpha\rho}Q_{\epsilon\gamma\rho}\\
	&+\mathcal{D}_\alpha T_{\gamma\delta\epsilon} Q^{\alpha\delta\rho}Q^{(\gamma\epsilon)}{}_\rho+4\mathcal{D}_{[\alpha} T_{\gamma]}{}^{\gamma\epsilon}Q^{\alpha\mu\rho}Q_{\epsilon\mu\rho}-\mathcal{D}_\mu T^\mu Q_{\kappa\nu\rho}Q^{\kappa\nu\rho}+\mathcal{D}_\mu T^\mu Q_\nu Q^\nu\\
	&-4\mathcal{D}_\mu T_\nu Q^{\nu}{}_{\sigma\rho}Q^{\sigma\mu\rho}-4\mathcal{D}_\alpha T^{(\gamma\epsilon)\alpha}Q_{\epsilon\mu\rho}Q^\mu{}_\gamma{}^\rho-2\mathcal{D}_\alpha T^{\alpha\delta\epsilon}Q_{\delta\mu}{}^{\rho}Q^\mu{}_{\epsilon\rho}\\
	&+2\mathcal{D}_\mu T^\mu Q_{\sigma\nu\rho}Q^{\nu\sigma\rho}+4\mathcal{D}_\alpha T_\epsilon Q^{(\alpha\epsilon)\mu}Q_\mu-4\mathcal{D}_\alpha T^{\gamma\alpha\epsilon}Q_{(\epsilon\gamma)\mu}Q^\mu-2\mathcal{D}_\mu T^\mu \Bar{Q}_\nu Q^\nu\\
	&+4\mathcal{D}_{[\alpha} T_{\gamma]}{}^{\alpha\epsilon}Q^{\mu\gamma}{}_{\epsilon}Q_\mu-\frac{1}{2}\mathcal{D}_\alpha T^{\gamma\delta\epsilon}Q_{\delta\epsilon\rho}Q^{\rho\alpha}{}_\gamma-\mathcal{D}_\alpha T^{\gamma\delta\epsilon}Q_{(\gamma\delta)\rho}Q^{\rho\alpha}{}_\epsilon\\
	&+2\mathcal{D}_\alpha T^{\gamma\delta\epsilon}Q_{[\delta\rho]}{}^\alpha Q^\rho{}_{\gamma\epsilon},
	\end{split}
	\end{align}
	with 27 terms. And $\mathcal{G}_{TQ\mathcal{D}Q}$ reads
	\begin{align}
	\begin{split}
	\mathcal{G}_{TQ\mathcal{D}Q}&=\frac{1}{2}\mathcal{D}_\gamma Q_{\delta\epsilon\eta}Q^{\tau\epsilon\eta}T^{\gamma\delta}{}_\tau +\mathcal{D}_\gamma Q_{\delta\epsilon\eta}Q^{\delta\epsilon\tau}T^{\gamma\eta}{}_\tau -\frac{1}{2}\mathcal{D}_\gamma Q_{\delta\epsilon\eta}Q^{\tau \delta\epsilon}T^{\gamma\eta}{}_\tau\\
	&+\mathcal{D}_\gamma Q^{\delta\gamma\eta}Q^\rho T_{\delta\eta\rho}+\mathcal{D}_\gamma Q_{\delta\epsilon\eta}Q^{[\epsilon\gamma]\tau}T^{\delta\eta}{}_\tau-\frac{3}{2}\mathcal{D}_\gamma Q_{\delta\epsilon\eta}Q^{\tau\gamma\epsilon}T^{\delta\eta}{}_\tau\\
	&+\frac{3}{2}\mathcal{D}_\gamma Q^{\delta\gamma\eta}Q_{\rho\eta\tau}T_\delta{}^{\rho\tau}-2\mathcal{D}_\gamma Q^{\gamma\epsilon\eta}Q^\rho T_{\epsilon\eta\rho}-\mathcal{D}_\gamma Q_{\delta\epsilon\eta}Q^{\delta\tau\gamma}T^{\epsilon\eta}{}_\tau+\mathcal{D}_\gamma Q^{\delta\gamma\eta}Q^\rho T_{\eta\delta\rho}\\
	&+\mathcal{D}_\gamma Q_{\delta\epsilon\eta}Q^{[\epsilon\gamma]\tau}T^{\eta\delta}{}_\tau+\frac{1}{2}\mathcal{D}_\gamma Q_{\delta\epsilon\eta}Q^{\tau\gamma\epsilon}T^{\eta\delta}{}_\tau +\frac{5}{2}\mathcal{D}_\gamma Q^{\delta\gamma\eta}Q_{\rho\delta\tau}T_\eta{}^{\rho\tau}\\
	&-\frac{3}{2}\mathcal{D}_\gamma Q^{\gamma\delta\eta}Q_{\rho\delta\tau}T_\eta{}^{\rho\tau}+2\mathcal{D}_\gamma Q^{[\gamma\epsilon]\eta}Q_{\epsilon}{}^{\rho\tau} T_{\rho\eta\tau}+2\mathcal{D}_\gamma Q^{[\delta\gamma]\eta}Q^\rho{}_{\delta\tau}T_{\rho\eta}{}^\tau\\
	&-\mathcal{D}_\mu \Bar{Q}^\mu Q_{\nu\rho\sigma}T^{\rho\nu\sigma}-\mathcal{D}_\mu \Bar{Q}^\mu Q_\nu T^\nu+4\mathcal{D}_\gamma Q^{[\gamma\epsilon]\eta}Q_{\epsilon\eta\rho}T^\rho-2\mathcal{D}_\gamma Q^{\delta\gamma\eta}Q_{\eta\delta\rho}T^\rho\\
	&+\mathcal{D}_\mu \Bar{Q}^\mu \Bar{Q}_\nu T^\nu+2\mathcal{D}_\gamma Q^{\delta\gamma\eta}Q_{\rho\delta\eta}T^\rho-\mathcal{D}_\gamma Q^{\gamma\epsilon\eta}Q_{\rho\epsilon\eta}T^\rho-\frac{1}{2}\mathcal{D}_\gamma Q_{\delta\epsilon\eta}Q^{\epsilon\eta\tau}T^{\gamma\delta}{}_\tau\\
	&-\frac{1}{2}\mathcal{D}_\mu Q_\nu Q_\rho T^{\mu\nu\rho}-\mathcal{D}_\mu Q_\nu Q_\rho{}^{\mu\sigma}T^{\nu\rho}{}_\sigma+2\mathcal{D}_\mu Q^\mu Q^{\sigma\rho}{}_{[\tau} T_{\rho]\sigma}{}^\tau+2\mathcal{D}_\mu Q_\nu Q^{(\mu\nu)\rho}T_\rho\\
	&-\mathcal{D}_\mu Q^\mu \Bar{Q}_\nu T^\nu-\mathcal{D}_\mu Q_\nu Q^{\rho\mu\nu}T_\rho-2\mathcal{D}_\mu Q_\nu Q_\rho{}^{\nu\sigma}T^{(\mu\rho)}{}_\sigma+\mathcal{D}_\mu Q_\nu Q_\rho{}^{\nu\sigma}T_\sigma{}^{\mu\rho}\\
	&-\mathcal{D}_\gamma Q_{\eta\delta}{}^\gamma Q_\rho{}^{\eta\tau}T_\tau{}^{\delta\rho}+\mathcal{D}_\gamma Q^{\gamma\epsilon\eta}Q^\rho{}_{\epsilon\tau}T^\tau{}_{\eta\rho}
	\end{split}
	\end{align}
	with 42 terms. And $\mathcal{G}_{TT\mathcal{D}Q}$ reads
	\begin{align}
	\begin{split}
	\mathcal{G}_{TT\mathcal{D}Q}&=-2\mathcal{D}_\alpha Q_{\gamma\delta\epsilon}T^{(\alpha\gamma)\rho}T^{\delta\epsilon}{}_\rho+\frac{1}{2}\mathcal{D}_\alpha Q_{\gamma\delta\epsilon}T^{\gamma\delta\rho}T^{\epsilon\alpha}{}_\rho +\frac{1}{2}\mathcal{D}_\alpha Q_{\gamma\delta\epsilon}T^{\delta\alpha\rho}T^{\epsilon\gamma}{}_\rho\\
	&+2\mathcal{D}_\alpha Q^{\gamma\alpha\epsilon}T_{\gamma\mu\rho}T_\epsilon{}^{\mu\rho}-\mathcal{D}_\alpha Q^{\alpha\delta\epsilon}T_{\delta\mu\rho}T_\epsilon{}^{\mu\rho}-\frac{3}{2}\mathcal{D}_\mu \bar{Q}^\mu T_{\nu\sigma\rho}T^{\nu\sigma\rho}\\
	&+4\mathcal{D}_\alpha Q^{\gamma\alpha\epsilon}T_{\mu\gamma}{}^{\rho}T^\mu{}_{\epsilon\rho}-2\mathcal{D}_\alpha Q^{\alpha\delta\epsilon}T_{\mu\delta}{}^\rho T^\mu{}_{\epsilon\rho}-\mathcal{D}_\mu \Bar{Q}^\mu T_{\sigma\nu\rho}T^{\nu\sigma\rho}+8\mathcal{D}_\alpha Q^{[\alpha\delta]\epsilon}T_{\delta\epsilon\mu}T^\mu \\
	&-4\mathcal{D}_\alpha Q^{\gamma\alpha\epsilon}T_{\epsilon\gamma\mu}T^\mu+2\mathcal{D}_\mu\Bar{Q}^\mu T_\nu T^\nu+2\mathcal{D}_\alpha Q_{\gamma\delta\epsilon} T^{\delta\epsilon\rho}T_\rho{}^{\alpha\gamma}+\mathcal{D}_{[\alpha} Q_{\gamma]\delta\epsilon}T^{\delta\alpha\rho}T_\rho{}^{\gamma\epsilon}\\
	&+\mathcal{D}_\alpha Q_{\gamma\delta\epsilon}T^{[\rho\alpha]\delta}T_\rho{}^{\gamma\epsilon}+\mathcal{D}_\alpha Q_{\gamma\delta\epsilon}T^{\alpha\delta\rho}T^{(\gamma\epsilon)}{}_\rho-\mathcal{D}_\alpha Q^\gamma T^{\alpha\mu\rho}T_{\gamma\mu\rho}\\
	&+\frac{3}{2}\mathcal{D}_\mu Q^\mu T_{\sigma\nu\rho}T^{\sigma\nu\rho}+\mathcal{D}_\mu Q^\mu T_{\sigma\nu\rho}T^{\nu\sigma\rho}-2\mathcal{D}_\mu Q^\mu T_\nu T^\nu +4\mathcal{D}_\alpha Q_\gamma T^{(\alpha\gamma)\mu}T_\mu\\
	&-2\mathcal{D}_\alpha Q_\gamma T^{\mu\alpha\gamma}T_\mu+4\mathcal{D}_\alpha Q^\gamma T^{\mu\alpha\rho} T_{(\mu\rho)\gamma}+4\mathcal{D}_\alpha Q^{\gamma\alpha\epsilon}T_{\mu\gamma}{}^{\rho}T_{\rho\epsilon}{}^\mu\\
	&-2\mathcal{D}_\alpha Q^{\alpha\delta\epsilon}T^\mu{}_{\delta\rho}T^\rho{}_{\epsilon\mu}-\frac{1}{2}\mathcal{D}_{\alpha}Q_{\gamma\delta\epsilon}T^{\gamma\delta\rho}T_\rho{}^{\alpha\epsilon},
	\end{split}
	\end{align}
	with 33 terms. Finally, the last term we need reads
	\begin{align}
	\begin{split}
	\mathcal{G}_{\mathcal{D}T\mathcal{D}Q}&=2\mathcal{D}_\nu Q_\sigma{}^{\mu}{}_{[\mu}\mathcal{D}_{\rho]} T^{\nu\sigma\rho}+2\mathcal{D}_\mu Q^{\sigma\rho}{}_{[\nu}\mathcal{D}_{\rho]} T^{\nu\mu}{}_{\sigma}+2\mathcal{D}_\mu Q^{\sigma\rho}{}_{[\nu}\mathcal{D}_{\rho]} T_\sigma{}^{\mu\nu}\\
	&+2\mathcal{D}^\mu Q_{\rho\sigma}{}^{[\nu}\mathcal{D}_\mu T^{\sigma]\rho}{}_\nu +\mathcal{D}_\mu Q^\mu{}_{\sigma\nu}\mathcal{D}_\rho T^{\sigma\nu\rho} +2\mathcal{D}_\mu Q^{\mu\nu}{}_{[\rho}\mathcal{D}_{\nu]} T^{\rho}+2\mathcal{D}_\mu Q^{\nu\mu}{}_{[\nu}\mathcal{D}_{\rho]} T^\rho\\
	&-\mathcal{D}_\mu Q^{\sigma\mu\nu}\mathcal{D}_\rho T_{\sigma\nu}{}^\rho -\mathcal{D}_\mu Q^{\nu\mu\rho}\mathcal{D}_{\rho} T_\nu-\mathcal{D}_\mu Q^{\sigma\mu\nu}\mathcal{D}_\rho T_\nu{}^{\sigma\rho}  
	\end{split}
	\end{align}
	with 16 terms. The result is summarized in Table \ref{genGBtable}, while the limits of vanishing nonmetricity and vanishing torsion are summarized in Tables \ref{teletable} and \ref{symtable}, respectively.

	%\begin{thebibliography}{99}
	%\bibliographystyle{acm}
 \bibliographystyle{ieeetr}
	\bibliography{references.bib}
	
	%\end{thebibliography}
	
\end{document}